\documentclass[11pt]{article}
\usepackage{fullpage}

\usepackage{microtype}

\usepackage[utf8]{inputenc}
\usepackage{tikz,rotating, color,graphicx, subcaption,url, comment}
\usetikzlibrary{matrix,arrows,automata,shapes,patterns}
\usepackage{amsfonts,amssymb,amsmath,amsthm}
\theoremstyle{definition}

\usepackage{makecell}
\usetikzlibrary{calc,positioning,angles,quotes}
\usepackage{multicol}

\usepackage{caption, hyperref}

\newtheorem{theorem}{Theorem}
\newtheorem{corollary}{Corollary}[theorem]
\newtheorem{lemma}[theorem]{Lemma}
\newtheorem{definition}{Definition}[section]

\newcommand{\toto}{xxx}
\newenvironment{proofT}{\noindent{\bf
		Proof }} {\hspace*{\fill}$\Box_{Theorem~\ref{\toto}}$\par\vspace{3mm}}
\newenvironment{proofL}{\noindent{\bf
		Proof }} {\hspace*{\fill}$\Box_{Lemma~\ref{\toto}}$\par\vspace{3mm}}

\newcommand{\vir}[1]{``#1''}

\usepackage{xargs}                      
\usepackage{xcolor}  
\usepackage[colorinlistoftodos,prependcaption,textsize=tiny]{todonotes}

\newcommandx{\YAG}[2][1=]{\todo[linecolor=black,backgroundcolor=teal!25,bordercolor=teal,#1]{\textcolor{teal}{\textbf{Yackolley}}\\#2}}
\newcommandx{\ST}[2][1=]{\todo[linecolor=black,backgroundcolor=red!25,bordercolor=red,#1]{\textcolor{teal}{\textbf{Sara}}\\#2}}
\newcommandx{\AD}[2][1=]{\todo[linecolor=blue,backgroundcolor=blue!25,bordercolor=blue,#1]{\textcolor{teal}{\textbf{Antonella}}\\#2}}
\newcommandx{\MP}[2][1=]{\todo[linecolor=green,backgroundcolor=green!25,bordercolor=green,#1]{\textcolor{teal}{\textbf{Maria}}\\#2}}

\usepackage[normalem]{ulem}
\newcommand\redout{\bgroup\markoverwith
	{\textcolor{red}{\rule[0.5ex]{2pt}{0.8pt}}}\ULon}

\makeatletter
\let\latex@@line\line
\def\line{\@ifnextchar(\latex@@line{\hbox to\hsize}}
\makeatother

\usepackage{amsthm}

\newcommand \nil{\textit{nil}}

\newcommand \Bool{\ensuremath{\textsf{Bool}}}
\newcommand \LLR{\ensuremath{\mathit{LLR}}}
\newcommand \PoLCR{\ensuremath{\mathit{PoLCR}}}
\newcommand \lockedBlock{\ensuremath{\mathit{lockedBlock}}}
\newcommand \TimeOutPropose{\ensuremath{\mathit{TimeOutPropose}}}
\newcommand \TimeOutPrevote{\ensuremath{\mathit{TimeOutPrevote}}}
\newcommand \proposalReceived{\ensuremath{\mathit{proposalReceived}}}
\newcommand \prevotesReceived{\ensuremath{\mathit{prevotesReceived}}}
\newcommand \precommitsReceived{\ensuremath{\mathit{precommitsReceived}}}
\newcommand \Block{\ensuremath{\mathit{Block}}}
\newcommand \MemPool{\ensuremath{\mathit{MemPool}}}
\newcommand \commitsReceived{\ensuremath{\mathit{commitsReceived}}}
\newcommand \toReward{\ensuremath{\mathit{toReward}}}
\newcommand \TimeOutCommit{\ensuremath{\mathit{TimeOutCommit}}}
\newcommand \Height{\ensuremath{\mathit{Height}}}

\newcommand\Pitem{%
	\addtocounter{enumi}{-1}%
	\renewcommand\theenumi{\arabic{enumi}'}%
	\item%
	\renewcommand\theenumi{\arabic{enumi}}%
}

\title{Correctness and Fairness of Tendermint-core  Blockchains}
\author{Yackolley Amoussou-Guenou$^{\ddagger,\star}$, Antonella Del Pozzo$^{\ddagger}$, \\Maria Potop-Butucaru$^\star$, Sara Tucci-Piergiovanni$^\ddagger$\\~\\
	$^\ddagger$CEA LIST, PC 174, Gif-sur-Yvette, 91191, France\\
	$^\star$Sorbonne Université, CNRS, Laboratoire d'Informatique de Paris 6, LIP6, Paris, France \\
}
\date{}
\usepackage{placeins}

\begin{document}
	
		\newcounter{linecounter}
		\newcommand{\linenumbering}{(\arabic{linecounter})}
		\renewcommand{\line}[1]{\refstepcounter{linecounter}
			\label{#1}
			\linenumbering}
		\newcommand{\resetline}{\setcounter{linecounter}{0}}

\maketitle

\begin{abstract}
Tendermint-core blockchains (e.g. Cosmos) are considered today one of the most viable alternatives for the highly energy consuming proof-of-work blockchains such as Bitcoin and Ethereum. Their particularity is that they
aim at offering strong consistency (no forks) in an open system combining two ingredients (i) a set of validators that generate blocks via a variant of Practical Byzantine Fault Tolerant (PBFT) consensus protocol and (ii) a selection strategy that dynamically selects nodes to be validators for the next block via a proof-of-stake mechanism. 
However, the exact assumptions on  the system model  under which Tendermint underlying algorithms are correct and the exact properties Tendermint verifies have never been formally analyzed. The contribution of this paper is two-fold. First, while formalizing Tendermint algorithms 
we precisely characterize the system model and the exact problem solved by Tendermint. We prove that  in eventual synchronous systems 
 a modified version of Tendermint solves (i) under additional assumptions, a variant of one-shot consensus for the validation of one single block and (ii) a variant of the repeated consensus problem for multiple blocks. These results hold even if the set of validators is hit by Byzantine failures, provided that for each one-shot consensus  instance  less than one third of the validators is Byzantine. Our second contribution relates to the fairness of the rewarding mechanism.  It is common knowledge that in permisionless blockchain systems  the main threat is the tragedy of commons that may yield the system to collapse if the rewarding mechanism is not adequate. Ad minimum the rewarding mechanism must be $fair$, i.e. distributing the rewards in proportion to the merit of participants.  We prove, for the first time in blockchain systems,  that in  repeated-consensus based blockchains  there exists an (eventual) fair rewarding mechanism if and only if the system is (eventual) synchronous.
We also show that the original Tendermint rewarding is not fair, however, a modification of the original protocol makes it eventually fair. 

 \end{abstract}
 
%
 
	\section{Introduction} \label{sec:intro}
Blockchain is today one of the most appealing technologies since its introduction in the Bitcoin White Paper \cite{bitcoinNakamoto} in 2008. 
  Blockchain systems, similar to P2P systems in the early 2000, take their roots in the non academical research.
After the releasing of the most popular blockchains  (e.g. Bitcoin \cite{bitcoinNakamoto} or Ethereum \cite{Ethereum}) with a specific focus on economical transactions, their huge potential  for various other applications ranging from notary to medical data recording became evident.  
In a nutshell, Blockchain systems maintain a continuously-growing history of ordered information, encapsulated in blocks. Blocks are linked to each other by relying on collision resistant hash functions, i.e., each block contains the hash of the previous block. The Blockchain itself is a distributed data structure replicated among different peers. In order to preserve the chain structure those peers need to agree on the next block to append in order to avoid forks. The most popular technique to decide which  block will be appended is  the \emph{proof-of-work} mechanism of Dwork and Naor \cite{DworkN92}. The block that will be appended to the blockchain is owned by the node (miner) having enough CPU power to solve first a crypto-puzzle. The only possible way to solve this puzzle is by repeated trials.
The major criticisms for the \emph{proof-of-work} approach are as follows: it is assumed that the honest miners hold a majority of the computational power, the generation of a block is energetically costly, which yield to the creation of mining pools and finally, multiple blockchains that coexist in the system due to accidental or intentional forks. 

Recently, the non academic research developed  alternative solutions to the proof-of-work technique such as \emph{proof-of-stake} (the power of block building is proportional to the  participant wealth),  \emph{proof-of-space} (similar to proof-of-work, instead of CPU power the prover has to provide the evidence of a certain amount of space) or \emph{proof-of-authority} (the power of block building is proportional to the amount of authority  owned in the system).  These alternatives received little attention  in the academic research.  Among all these alternatives \emph{proof-of-stake} protocols and in particular those using variants of \emph{Practical Byzantine Fault-Tolerant}  consensus \cite{pbft} became recently popular not only for in-chain transaction systems but also in systems that provide cross-chain transactions. 
Tendermint \cite{Tendermint,buchman-thesis2016,know-manuscript2014,malkhi-blog} was the  first in this line of research having the merit to link the  \emph{Practical Byzantine Fault-Tolerant}  consensus to the proof-of-stake technique and to propose  a blockchain where a dynamic set of validators (subset of the participants) decide on the next block to be appended to the blockchain. Although, the  correctness of the original Tendermint protocol \cite{Tendermint,buchman-thesis2016,know-manuscript2014}  has never been formally analyzed from the distributed computing perspective, it or slightly modified variants became recently the core of several popular systems such as 
Cosmos \cite{cosmos} for cross-chain transactions.  

 In this paper we analyse the correctness of the original Tendermint agreement protocol as it was described in \cite{Tendermint,buchman-thesis2016,know-manuscript2014} and discussed in \cite{malkhi-blog,HM16}. The code of this protocol is available in \cite{github-tendermint}. One of our  fundamental results proved in this paper is as follows:
 
\begin{centering}
	\textit{In an eventual synchronous system, a slightly modified variant of the original Tendermint protocol implements the one-shot and repeated consensus,  provided that (i) the number of Byzantine validators, $f$, is $f<n/3$ where $n$ is the number of validators participating in each single one-shot consensus instance and (ii)  eventually a proposed value will be accepted by  at least $2n/3+1$ processes (Theorem \ref{t:consensus} and Theorem \ref{t:repConsensus}).}
\end{centering}

More in detail, we prove that the original Tendermint verifies the consensus termination with a small twist in the algorithm (a refinement of the timeout) and with the additional assumption stating   that there exists eventually a proposer such that its proposed value will be accepted, or voted, by more than two-third of validators. 


We are further interested in the \emph{fairness}  of  Tendermint-core  blockchains
because without a glimpse of fairness in the way rewards are distributed, these blockchains may collapse. 
Our fairness study, in line with Francez definition of fairness \cite{francez86},  generally defines the fairness of protocols based on voting committees (e.g. Byzcoin\cite{BizCoin}, PeerCensus\cite{PeerCensus}, RedBelly \cite{redbelly17}, SBFT \cite{sbft18} and Hyperledger Fabric \cite{hyperledger} etc), by the fairness of their \emph{selection mechanism}  and the fairness of their \emph{reward mechanism}. 
The selection mechanism is in charge of selecting the subset of processes that will participate to the agreement on the next block to be appended to the blockchain, while
the reward mechanism defines the way the rewards are distributed among  processes that participate in the agreement. 
The analysis of the reward mechanism  allowed to establish our second fundamental result with respect to the fairness of  repeated-consensus blockchains  as follows:

\begin{centering}
\textit{There exists a(n) (eventual) fair reward mechanism for  repeated-consensus blockchains   if and only if the system is (eventual) synchronous (Theorem \ref{t:eventualFairness}).}
 \end{centering}
 
Moreover, we show that even in an eventual synchronous setting, the original Tendermint protocol is not eventually fair, however with a small twist in the way delays and commit messages are handled it becomes eventually fair.
 
Note that our work is the first to analyze the fairness of protocols based on voting committees elected by  selection mechanisms as \emph{proof-of-stake}. %
 
 The rest of the paper is organized as follows.  Related works are discussed in Section  \ref{sec:relatedwork}. Section \ref{sec:model} defines the model and the formal specifications of one-shot and repeated consensus. Section \ref{sec:formalization} formalizes the original Tendermint protocol through pseudo-code and proves the correctness of the One-Shot Consensus anf the Repeated Consensus algorithms.
In Section \ref{sec:scenario} we present a full descriptions of the counter-example that motivates the modification of the original algorithm and the additioinal assumptions needed for the correctness.
Section \ref{sec:fairness} discusses the necessary and sufficient conditions for a protocol based on repeated consensus to achieve a fair rewarding.

		\section{Related Work} \label{sec:relatedwork}

Interestingly, only recently distributed computing academic scholars focus their attention on the  theoretical aspects of blockchains motivated mainly by the intriguing claim of popular blockchains, as Bitcoin and Ethereum,  that they implement consensus in an asynchronous  dynamic open system. This claim is refuted by the famous impossibility result in  distributing computing \cite{FLP}. 
 
In distributed systems, the theoretical studies of   \emph{proof-of-work} based blockchains have been  pioneered by Garay et \emph{al}~\cite{GarayKL15}.   Garay \emph{et al.} decorticate the pseudo-code of Bitcoin  and  analyse its agreement aspects considering a synchronous round-based communication model. 
This study  has been extended by Pass \emph{et al.}~\cite{PSS17}  to round based systems where messages sent in a round can be received later.  
In order to overcome the drawbacks of Bitcoin, \cite{Bitcoin-NG} proposes a mix between proof-of-work blockchains and proof-of-work free blockchains referred as Bitcoin-NG. 
Bitcoin-NG inherits the drawbacks of Bitcoin: costly proof-of-work process, forks, no guarantee that a leader in an epoch is unique, no guarantee that the leader does not change the history at will if it is corrupted.

On another line of research Decker \emph{et al.} \cite{PeerCensus}   propose the PeerCensus system that targets linearizability of transactions. PeerCensus combines   the proof-of-work blockchain and the classical results in Practical Byzantine Fault Tolerant agreement area. PeerCensus suffers  the same drawbacks as Bitcoin and Byzcoin against dynamic adversaries. 

Byzcoin \cite{BizCoin} builds on top of \emph{Practical Byzantine Fault-Tolerant}  consensus \cite{pbft}  enhanced with a scalable collective signing process.   \cite{BizCoin} is based on a leader-based consensus over a group of members chosen by a proof-of-membership mechanism. 
When a miner succeeds to mine a block, it gains a membership share, and the miners with the highest shares are part of the fixed size voting member set.
In the same spirit,   SBFT \cite{sbft18} and Hyperledger Fabric \cite{hyperledger}  build on top of \cite{pbft}. 

In order to avoid some of the previously cited problems,  Micali~\cite{micali2016} introduced (further extended in \cite{BPS16,micali2017a})
\emph{sortition} based blockchains, where the proof-of-work mechanism is completely replaced by  a probabilistic ingredient.  

 The only academic work that addresses the consensus in \emph{proof-of-stake} based blockchains is authored by Daian \emph{et al.} \cite{DPS16a}, which proposes a protocol for weakly synchronous networks. The execution of the protocol is organized in epochs. Similar to Bitcoin-NG  \cite{Bitcoin-NG}  in each epoch a different committee is elected and inside the elected committee a leader will be chosen. The leader is allowed to extend the new blockchain.  The protocol is validated via simulations and only partial proofs of correctness are provided.  Ouroboros \cite{KRDO17} proposes a sortition based proof-of-stake protocol and addresses mainly the security aspects of the proposed protocol. Red Belly \cite{redbelly17} focuses on  consortium blockchains,  where  only a predefined subset of processes are allowed to append blocks, and proposes a Byzantine consensus protocol.

 Interestingly, none of the previous academic studies made the connection between the repeated consensus specification \cite{NDGK95,DR03,DDFPT08} and the repeated agreement process in blockchain systems. Moreover,  in terms of fairness of rewards, no academic study has been conducted related to blockchains based on repeated  consensus.

 The closest works in blockchain systems to our fairness study (however very different in its scope) study the \emph{chain-quality}.
In \cite{GarayKL15}, Garay \emph{et al.} define the notion of \emph{chain-quality} as  the proportion of blocks mined by honest miners in any given window and study the conditions under which the ratio of blocks in the chain mined by malicious players over the total number of blocks in a given window is bounded.
Kiayias \emph{et al.} in \cite{KRDO17} propose Ouroboros \cite{KRDO17} and  also analyse the chain-quality property.
Pass \emph{et al.} address in \cite{PS17}   one of the vulnerabilities of Bitcoin studied formally in Eyal and Sirer \cite{ES14}. In \cite{ES14} the authors  prove that if the adversary controls a coalition of miners holding even a  minority of the computational power, this coalition can gain twice its share. Fruitchain \cite{PS17} overcomes this problem by ensuring that no coalition controlling
less than a majority of the computational power can gain more than a factor $1+3\delta$ by not respecting  the protocol, where $\delta$ is a parameter of the protocol. 
In \cite{Eyal15}, Eyal analyses the consequences of attacks in systems where pools of miners can infiltrate each other and shows that in such systems there is an equilibrium where all pools earn less than if there were no attack.
In \cite{GW18}, Guerraoui and Wang study the effect of message propagation delays in Bitcoin and show that, in a system of two miners, a miner can take advantage of the delays and be rewarded exponentially more than its expectation.
In \cite{GDT17}, G{\"u}rcan \emph{et al.} study the fairness from the point of view of users that do not participate to the mining.
A similar work is done by Herlihy and Moir in \cite{HM16} where the authors study the users fairness and consider as an example the original Tendermint \cite{Tendermint,buchman-thesis2016,know-manuscript2014}. The authors discussed how processes with malicious behaviour can violate fairness by choosing transactions, then they propose
modifications to the original Tendermint to make those violations detectable and accountable.



	\section{System model and Problem Definition} \label{sec:model}

The system is composed of an infinite set $\Pi$ of asynchronous sequential processes, namely $\Pi = \{p_1, \dots\}$;
$i$ is called the \textit{index} of $p_i$. \textit{Asynchronous} means that each process proceeds at it own speed,
which can vary with time and remains unknown to the other processes. \textit{Sequential} means that a process executes one step at a time.
This does not prevent it from executing several threads with an appropriate multiplexing. As local processing time are negligible with respect to message transfer delays, they are considered as being equal to zero.

\textbf{Arrival model.}
We assume  a \textit{finite arrival model} \cite{aguilera2004}, i.e.  the system has  infinitely many processes but each run has only finitely many.   The size of the set $\Pi_{\rho} \subset \Pi$  of processes that participate  in each system run is not a priori-known. 
We also consider a finite subset $V \subseteq \Pi_\rho$  of validators. The set $V$ may change during any system run and its size $n$ is a-priori known. A process is promoted in $V$ based on a so-called merit parameter, which can model for instance its stake in proof-of-stake blockchains. Note that in the current Tendermint implementation, it is a separate module included in the Cosmos project \cite{cosmos} that is  in charge of implementing the selection of $V$.

\textbf{Communication network.} The processes communicate by exchanging messages through an eventually synchronous  network \cite{DLS88}. \textit{Eventually Synchronous} means that after a finite unknown time  $\tau$ there is a bound $\delta$ on the message transfer delay.

\textbf{Failure model.} There is no bound on processes that can exhibit a Byzantine behaviour \cite{RA80} in the system, but up to $f$ validators can exhibit a Byzantine behaviour at each point of the execution. A Byzantine process is a process that behaves arbitrarily: it can crash, fail to send or receive messages, send arbitrary messages, start in an arbitrary state, perform arbitrary state transition, etc. Byzantine processes can control the network by modifying the order in which messages are received, but they cannot postpone forever message receptions.
Moreover, Byzantine processes can collude to \vir{pollute} the computation (e.g., by sending messages with the same content, while they should send messages with distinct content if they were non-faulty).
A process (or validator) that exhibits a Byzantine behaviour is called \textit{faulty}. Otherwise, it is \textit{non-faulty} or \textit{correct} or \textit{honest}.
To be able to solve the consensus problem, we assume that $f < n/3$. 

\textbf{Communication primitives.} 
In the following we assume the presence of a broadcast primitive. A process $p_i$ broadcasts a message by invoking the primitive $\textsf{broadcast} (\langle \textit{TAG},m \rangle)$, where $\textit{TAG}$ is the type of the message, and $m$ its content. To simplify the presentation, it is assumed that a process can send messages to itself. The primitive $\textsf{broadcast} ()$ is a best effort broadcast, which means that when a correct process broadcasts a value, eventually all the correct processes deliver it.
A process $p_i$ receives a message by executing the primitive $\textsf{delivery}()$.
Messages are created with a digital signature, and we assume that digital signatures cannot be forged. When a process $p_i$ delivers a message, it knows the process $p_j$ that created the message.

Let us note that the assumed broadcast primitive in an open dynamic network can be implemented through $gossiping$, i.e. each process sends the message to current neighbors in the underlying dynamic network graph. In these settings the finite arrival model is a necessary condition for the system to show eventual synchrony. Intuitively, a finite arrival implies that message losses due to topology changes are bounded, so that the propagation delay of a message between two processes not directly connected  can be bounded {\cite{BBTR2007}.}

\textbf{Problem definition.}
In this paper we analyze the correctness of Tendermint protocol against two abstractions in distributed systems: consensus and repeated consensus defined formally as follows.

\begin{definition}[One-Shot Consensus]
	We say that an algorithm implements One-Shot Consensus if and only if it satisfies the following properties:
	\begin{itemize}
		\item \textbf{Termination.} Every correct process eventually decides some value.
		\item \textbf{Integrity.} No correct process decides twice.
		\item \textbf{Agreement.} If there is a correct process that decides a value $B$, then eventually all the correct processes decide $B$.
		\item \textbf{Validity\cite{redbelly17}.} A decided value is valid, it satisfies the predefined predicate denoted $\textsf{isValid}()$.
	\end{itemize}
	
\end{definition}	
	
The concept of multi-consensus is presented in \cite{NDGK95}, where the authors assume that only the faulty processes can postpone the decision of correct processes. In addition, the consensus is made a finite number of times. The long-lived consensus presented in \cite{DR03} studies the consensus  when the inputs are  changing over the time, their specification aims at studying in which condition the decisions of correct process do not change over time.
None of these specifications is appropriate for blockchain systems. 
In \cite{DDFPT08}, Delporte-Gallet \emph{et al.} defined the Repeated Consensus as an infinite sequence of One-Shot Consensus instances, where the inputs values may be completely different from one instance  to another, but where all the correct processes have the same infinite sequence of decisions.
We consider a variant of the repeated consensus problem as defined in \cite{DDFPT08}.
The main difference is that we do not predicate on the faulty processes.
Each correct process outputs an infinite sequence of decisions. We call that sequence the \emph{output} of the process. 
	
\begin{definition}[Repeated Consensus]
	An algorithm implements a repeated consensus if and only if it satisfies the following properties:
	\begin{itemize}
		\item \textbf{Termination.} Every correct process has an infinite output.
		\item \textbf{Agreement.} If the $i^{th}$ value of the output of a correct process is $B$, then $B$ is the $i^{th}$ value of the output of any other correct process.
		\item \textbf{Validity.} Each value in the output of any correct process is valid, it satisfies the predefined predicate denoted $\textsf{isValid}()$.
	\end{itemize}
\end{definition}
	
	\section{ Tendermint Formalization}\label{sec:formalization}
		\subsection{Informal description of Tendermint and its blockchain}\label{ssec:description}

Tendermint protocol  \cite{Tendermint,buchman-thesis2016} aims at building a blockchain without forks relying on a variant of PBFT consensus. 
When building the blockchain, a subset of fixed size $n$ of processes called validators should agree on the next block to append to the blockchain. The set of validators is deterministically determined by the current content of the blockchain, referred as the history. We note that this subset may change once a block is appended.
The mechanism to choose the validators from a given history is further referred as \emph{selection mechanism}. Note that in the current Tendermint implementation, it is a separate module included in the Cosmos project \cite{cosmos} that is  in charge of implementing the selection mechanism.  Intuitively, such mechanism should be based on the proof-of-stake approach but its actual implementation is currently left open.  Let us recall that Tendermint agreement protocol relies on the assumption that the selection mechanism, for each block, selects up to $f$ Byzantine processes from the current history.

The first block of Tendermint blockchain, called the \emph{genesis block}, is at \emph{height} $0$. The \emph{height} of a block is the distance that separates that block to the genesis block.
Each block contains: (i) a \emph{Header} which contains a pointer to the previous block and the height of the block, (ii) the \emph{Data} which is a list of transactions, and (iii) a set \emph{LastCommit} which is the set of validators that signed on the previous block. Except the first block, each block refers to the previous block in the chain.
Given a current height of Tendermint blockchain, a total ordered set of  validators $V$ is selected to add a new block. The validators start a \emph{One-Shot Consensus algorithm}. The first validator creates and proposes a block $B$, then if more than $2n/3$ of the validators accept $B$, $B$ will be appended as the next block, otherwise the next validator proposes a block, and the mechanism is repeated until more than $2n/3$ of the validators accept a block.  
For each height of Tendermint blockchain, the mechanism to append a new block is the same, only the set of validators may change. Therefore, Tendermint  applies a \emph{Repeated Consensus algorithm} to build a blockchain, and at each height, it relies on a \emph{One-Shot Consensus algorithm} to decide the block to be appended. 

Although  the choice of validators is managed by a separate module (see Cosmos project \cite{cosmos})   the rewards for the validators that contributed to the block at  some specific height $H$ are determined during the construction of the block at height $H+1$.
The validators for $H$ that get a reward for $H$ are the ones that validators for $H+1$ \vir{saw} when proposing a block.
This mechanism can be unfair, since some validator for $H$ may be slow, and its messages may not reach the validators involved in $H+1$, implying that it may not get the rewards it deserved. 

		
		\subsection{Tendermint One-Shot Consensus algorithm}\label{ssec:algoConsensus}

\begin{figure*}[h!]
\resizebox{\textwidth}{!}{
\fbox{
	\centering
	\begin{minipage}{0.4\textwidth}
		\scriptsize
		\begin{tabbing}
			aaaA\=aA\=aA\=aA\=aA\=aaaA\kill
			{\bf Function} {\sf consensus}$(H, \Pi, \text{signature})$; \%One-Shot Consensus for the super-round $H$ with the set $\Pi$ of processes\%\\ \\
			
			{\bf Init}:\\
				\line{P-01} \> $r \leftarrow 0$; $\LLR_i \leftarrow -1$; $\PoLCR_i \leftarrow \bot$; $\lockedBlock_i \leftarrow \nil$; $B \leftarrow \bot ;$\\
				\line{} \> $\TimeOutPropose \leftarrow \Delta_{\text{Propose}}$; $\TimeOutPrevote \leftarrow \Delta_{\text{Prevote}}$;\\
				\line{} \> $\proposalReceived^{H,r}_i \leftarrow \bot$; $\prevotesReceived^{H,r}_i \leftarrow \bot$; $\precommitsReceived^{H,r}_i \leftarrow \bot$;\\
			~-----------------------------------------------------------------------------------------------------------------------------------------------\\
			{\bf while} ${\sf (true)}$ {\bf do}\\
				\line{} \> $r \leftarrow r + 1$; $\PoLCR_i \leftarrow \bot;$\\
				
				~------------------------------------------------------------ Propose step $r$ ------------------------------------------------------------\\
				\line{Prop-begin} \> {\bf if} $(p_i == {\sf proposer}(H,r))$ {\bf then}\\
					\line{} \>\> {\bf if} $(\LLR_i \neq -1)$ {\bf then} $\PoLCR_i \leftarrow \LLR_i$; $B \leftarrow \lockedBlock_i$;\\
					\line{} \>\>\> {\bf else} $B \leftarrow {\sf createNewBlock}(\text{signature})$;\\
					\line{} \>\> {\bf endif}\\
					\line{Prop-bc1} \>\> {\bf trigger} {\sf broadcast} $\langle \texttt{PROPOSE}, (B,H,r, \PoLCR_i)_i \rangle$;\\
				\line{} \> \ {\bf else}\\
					\line{} \>\> {\sf set} $timerProposer$ {\sf to} $\TimeOutPropose$;\\
					\line{Prop-wait} \>\> {\bf wait until} (($timerProposer$ expired$) \lor (\proposalReceived^{H,r'}_i \neq \bot))$;\\
					\line{Prop-01} \>\> {\bf if} $((timerProposer$ expired$) \land (\proposalReceived^{H,r'}_i == \bot) )$ {\bf then}\\ 
						\line{} \>\>\> $\TimeOutPropose \leftarrow \TimeOutPropose + 1$;\\
					\line{Prop-02} \>\> {\bf endif}\\	
				\line{Prop-end} \> {\bf endif}\\	
				~------------------------------------------------------------ Prevote step $r$ ------------------------------------------------------------\\
				\line{PV-begin} \> {\bf if} $((\PoLCR_i \neq \bot) \land (\LLR_i \neq -1) \land (\LLR_i < \PoLCR_i < r) )$ {\bf then}\\
						\line{} \>\> {\bf wait until}  $|\prevotesReceived^{H,\PoLCR}_i| > 2n/3$;\\
						\line{PV-unlock} \>\> {\bf if} $(\exists B' : ({\sf is23Maj} (B',\prevotesReceived^{H,\PoLCR_i}_i)) \land (B' \neq \lockedBlock_i))$ {\bf then }
							$\lockedBlock_i \leftarrow \nil$;  {\bf endif}\\
				\line{} \> {\bf endif}\\	
				\line{PV-01} \> {\bf if} ($\lockedBlock_i \neq \nil)$ {\bf then}  {\bf trigger} {\sf broadcast} $\langle \texttt{PREVOTE}, (\lockedBlock_i,H,r)_i\rangle$;\\
				\line{PV-valid01} \> \ {\bf else} {\bf if} ({\sf isValid}$(\proposalReceived^{H,r}_i))$ {\bf then}  {\bf trigger} {\sf broadcast} $\langle \texttt{PREVOTE}, ( \proposalReceived^{H,r}_i,H,r)_i\rangle$; {\bf endif}\\	
				\line{} \> \ {\bf else} {\bf trigger} {\sf broadcast} $\langle \texttt{PREVOTE}, ( \nil,H,r)_i \rangle$;\\
				\line{PV-02} \> {\bf endif}\\	
				\line{PV-wait} \> {\bf wait until} $(({\sf is23Maj} (\nil,\prevotesReceived^{H,r}_i)) \lor (\exists B'':({\sf is23Maj} (B'',\prevotesReceived^{H,r}_i))) \lor$ \\ \>\>\>\>$(|\prevotesReceived^{H,r}_i| > 2n/3)$); \%Delivery of any 2n/3 prevotes for the round $r$\%\\
				
				\line{PV-timer} \> {\bf if} $(\lnot({\sf is23Maj} (\nil,\prevotesReceived^{H,r}_i)) \land \lnot(\exists B'':({\sf is23Maj} (B'',\prevotesReceived^{H,r}_i))))$ {\bf then}\\
					\line{} \>\> {\bf set} $timerPrevote$ {\bf to} $\TimeOutPrevote$;\\
					\line{} \>\> {\bf wait until} ($timerPrevote$ expired);\\
					\line{PV-end} \>\> {\bf if} $(timerPrevote$ expired$)$ {\bf then} $\TimeOutPrevote \leftarrow \TimeOutPrevote + 1$; {\bf endif}\\	
					
				~------------------------------------------------------------ Precommit step $r$ --------------------------------------------------------\\
				\line{PC-01} \> {\bf if} $(\exists B' : ({\sf is23Maj} (B',\prevotesReceived^{H,r}_i)))$ {\bf then}\\
					\line{PC-lock} \>\> $\lockedBlock_i \leftarrow B'$;\\
					\line{PC-02} \>\> {\bf trigger} {\sf broadcast} $\langle \texttt{PRECOMMIT}, (B',H,r)_i \rangle$;\\
					\line{} \>\> $\LLR_i \leftarrow r$;\\
				 \line{PC-03} \> \ {\bf else if} $({\sf is23Maj} (nil,\prevotesReceived^{H,r}_i))$ {\bf then}\\
				 	\line{} \>\> $\lockedBlock_i \leftarrow \nil$; $\LLR_i \leftarrow -1$;\\
					\line{PC-04} \>\> {\bf trigger} {\sf broadcast} $\langle \texttt{PRECOMMIT}, (\nil,H,r)_i \rangle$;\\
				 \line{} \> \ {\bf endif}\\		
				 \line{PC-05} \> \ {\bf else}  {\bf trigger} {\sf broadcast} $\langle \texttt{PRECOMMIT}, (\nil,H,r)_i \rangle)$;\\	
				 \line{} \> {\bf endif}\\	
				 \line{PC-wait} \>{\bf wait until} $(({\sf is23Maj} (\nil,\prevotesReceived^{H,r}_i)) \lor (|\precommitsReceived^{H,r}_i| > 2n/3)$)\\
				 {\bf endwhile} \\
		\end{tabbing}
		\normalsize
\end{minipage}%

	}
	}
		\caption{First part  of Tendermint One-shot Consensus algorithm at correct process $p_i$.}
\end{figure*}

\begin{figure*}[h!]\ContinuedFloat
\centering
\fbox{
	\begin{minipage}{0.4\textwidth}
		\scriptsize
		\begin{tabbing}
			aaaA\=aA\=aA\=aA\=aA\=aaaA\kill
			
			 {\bf upon event} {\sf delivery} $\langle \texttt{PROPOSE}, (B',H,r', \PoLCR_j)_j \rangle$:\\
			 	\line{} \>  {\bf if} $(\proposalReceived^{H,r'}_i = \bot)$ {\bf then}\\
					\line{del-Prop01} \>\> $\proposalReceived^{H,r'}_i \leftarrow (B',H,r')_j$;\\
					\line{} \>\> $ \PoLCR_i \leftarrow  \PoLCR_j$;\\
					\line{Prop-bc} \>\> {\bf trigger} {\sf broadcast} $\langle \texttt{PROPOSE}, (B',H,r', \PoLCR_j)_j \rangle$;\\
				\line{} \> {\bf endif}\\
			
			~------------------------------------------------------------------------------------------------------------\\
			 {\bf upon event} {\sf delivery} $\langle \texttt{PREVOTE}, ( B',H,r', \LLR)_j \rangle$:\\
			 	\line{} \>  {\bf if} $(( B',H,r', \LLR)_j \notin \prevotesReceived^{H,r'}_i)$ {\bf then}\\
					\line{PV-del} \>\> $\prevotesReceived^{H,r'}_i \leftarrow \prevotesReceived^{H,r'}_i \cup ( B',H,r', \LLR)_j$;\\
					\line{PV-bc} \>\> {\bf trigger} {\sf broadcast} $\langle \texttt{PREVOTE}, ( B',H,r', \LLR)_j \rangle$;\\
					\line{} \>\> {\bf if}  $((r < r')$ {\bf and} $(|\prevotesReceived^{H,r'}_i| >2/3))$ {\bf then}\\ 
						\line{} \>\>\> $r \leftarrow r'$;\\
						\line{} \>\>\> {\bf goto} Prevote step $r$;\\
					\line{} \>\> {\bf endif}\\
				\line{} \> {\bf endif}\\
			~------------------------------------------------------------------------------------------------------------\\
			 {\bf upon event} {\sf delivery} $\langle \texttt{PRECOMMIT}, ( B',H,r')_j\rangle$:\\
			 	\line{} \>  {\bf if} $(( B',H,r')_j \notin \precommitsReceived^{H,r'}_i)$ {\bf then}\\
					\line{} \>\> $\precommitsReceived^{H,r'}_i \leftarrow \precommitsReceived^{H,r'}_i \cup ( B',H,r')_j$;\\
					\line{PC-bc} \>\> {\bf trigger} {\sf broadcast} $\langle \texttt{PRECOMMIT},( B',H,r')_j\rangle$;\\
					\line{PC-exit01} \>\> {\bf if}  $((r < r')$ {\bf and} $(|\precommitsReceived^{H,r'}_i| > 2/3))$ {\bf then}\\ 
						\line{} \>\>\> $r \leftarrow r'$;\\
						\line{} \>\>\> {\bf goto} Precommit step $r$;\\
					\line{PC-exit02} \>\> {\bf endif}\\
				\line{} \> {\bf endif}\\
			~------------------------------------------------------------------------------------------------------------\\
			 {\bf when}  $(\exists B': {\sf is23Maj} (B',\precommitsReceived^{H,r'}_i))$:\\
				\line{return} \> {\bf return} B';\%Terminate the consensus for the super-round $H$ by deciding $B'$\%\\

		\end{tabbing}
		\normalsize
	\end{minipage}%

	}		
	\caption{Second part  of Tendermint One-shot Consensus algorithm at correct process $p_i$.}
	\label{fig:code}
\end{figure*}

%
	Tendermint One-Shot Consensus algorithm is a round-based algorithm used to decide on the next block for a given height $H$.
In each \textit{round} there is a different proposer that proposes a block to the validators that try to decide on that block. A round consists of three steps: (i) the \textit{Propose step}, the proposer of the round broadcasts a proposal for a block; (ii) the \textit{Prevote step}, validators broadcast their prevotes depending on the proposal they delivered during the previous step; and (iii) the \textit{Precommit step}, validators broadcast their precommits depending on the occurrences of prevotes for the same block they delivered during the previous step. To preserve liveness, steps have a timeout associated, so that each validator moves from one step to another either if the timeout expires or if it delivers enough messages of a particular typology.
When $p_i$ broadcasts a message $(\langle \textit{TAG},m \rangle)$, $m$ contains a block $B$ along with other information. We say that  $p_i$ prevotes (resp. precommits) on $B$ if $\textit{TAG}=\texttt{PREVOTE}$ (resp. $\textit{TAG}$=$\texttt{PRECOMMIT}$).  In Figure \ref{fig:stateMachine} is depicted the state machine for Tendermint Consensus. 
%


\begin{figure}[h!]
	\centering
	\resizebox{.8\linewidth}{!}{
		\begin{tikzpicture}[thick,scale=0.8, every node/.style={scale=0.90}]
		\tikzstyle{state}=[draw, circle, minimum size=2cm]
		\tikzstyle{block}=[draw,rectangle]
			\node[state] (new) at (4,-3.5)  [align=center] {New Height \\ (H{=H+1})};
			\node[state] (propose) at (5,0)  [align=center] {Propose \\ (H,r{=r+1})};
			
			\node[state] (prevote) at (10,0) [align=center] {Prevote \\ (H,r)};
			\node[state] (precommit) at (15,0) [align=center] {Precommit \\ (H,r)};
			\node[state] (commit) at (16,-3.5) [align=center] {Commit \\ (H)};
			
			\node[block, below = .3cm of propose] (proposeCode) [align=center] {lines \ref{Prop-begin} - \ref{Prop-end} };
			\node[block, below = .3cm of prevote] (prevoteCode) [align=center] {lines \ref{PV-begin} - \ref{PV-end}};
			\node[block, below = .3cm of precommit] (precommitCode) [align=center] {lines \ref{PC-01} - \ref{PC-wait}};
			
			\draw 
			(new) edge[bend left, above, ->] (propose)
			
			(propose) edge[->] node[align=left, yshift=1.6cm]{\scriptsize  \TimeOutPropose expired $\vee$ \\ \scriptsize  Propose(B,PolCR) delivered} (prevote)
			
			(prevote) edge[->] node[align=left, yshift=2cm] {\scriptsize More than $2n/3$ Prevote(B) $\vee$ \\ \scriptsize More than $2n/3$ Prevote(\nil) $\vee$ \\ \scriptsize \TimeOutPrevote expired $\vee$ \\ \scriptsize More than $2n/3$ of any Prevote(-)} (precommit)
			
			(precommit) edge[bend left, above, ->] node[align=left, xshift=1.5cm] {\scriptsize More than \\ \scriptsize $2n/3$ Precommit(B)} (commit)
			
			(precommit) edge[bend right=100, above, ->] node[align=left, xshift=6cm,yshift=-1cm]{\scriptsize More than $2n/3$ Prevote(\nil)  $\vee$ \\ \scriptsize TimeOutPrecommit expired $\vee$ \\ \scriptsize More than $2n/3$ of any Precommit(-) }(propose)

			(commit) edge[->] (new)

			
	%
	%
	%
	;
			
		\end{tikzpicture}
	}
	\caption{State Machine for Tendermint One-Shot algorithm described in Figure \ref{fig:code}. For ease of readability, common exit conditions are represented.}\label{fig:stateMachine}
\end{figure}
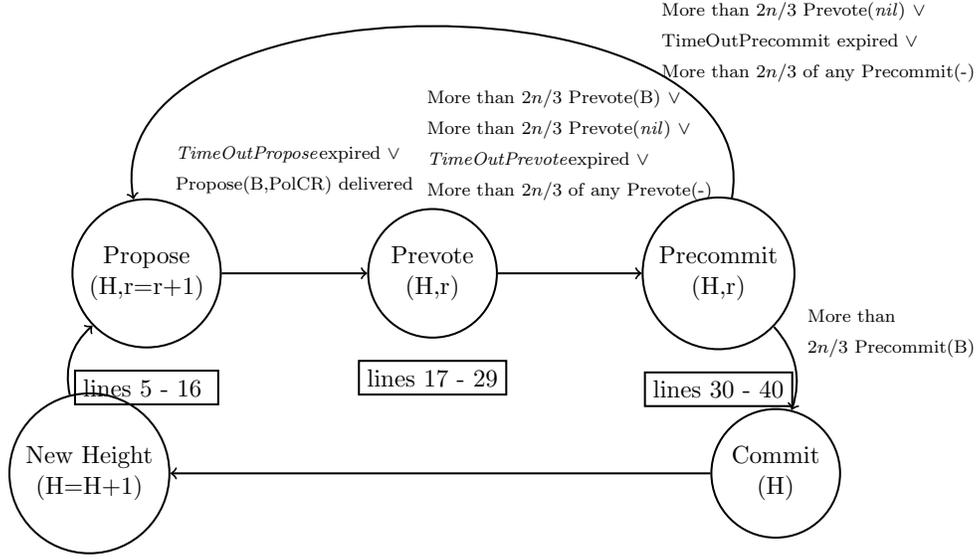

\begin{figure}[t!]
	\centering
	\resizebox{.75\linewidth}{!}{
		\begin{tikzpicture}[thick,scale=0.8, every node/.style={scale=0.90}]
			\tikzstyle{every state}=[draw, circle, minimum size=1.7cm,node distance=80mm]
			\tikzset{initial text={}}
			\node[initial above,state] (lock) [align=center]{$\exists B:$ \\ lock(B)};
			\node[state, right of = lock] (unlock) {lock(\nil)};
			
			\draw [->] (lock) edge[bend right] node[below,align=center]{\scriptsize More than $2n/3$ Prevote(\nil) $\lor$ \\ \scriptsize (\texttt{PROPOSE}(B',PolCR) delivered $\land$ \textsf{isValid}(B') $\land$ PolCR $\neq \bot) \lor$ \\ \scriptsize Commit} (unlock);
			
			\draw [->] (lock) edge[loop left] node[above,rotate=90,align=center]{\scriptsize $\nexists B':$ More than $2n/3$ Prevote(B') $\lor$ \\ \scriptsize $\exists B:$ More than $2n/3$ Prevote(B) } (lock);
			
			\draw [->] (unlock) edge[bend right] node[above]{\scriptsize More than $2n/3$ Prevote(B)}(lock);
			
			\draw [->] (unlock) edge[loop right] node[above, rotate=270]{\scriptsize $\nexists B:$ More than $2n/3$ Prevote(B)} (unlock);
		\end{tikzpicture}
	}
	\caption{State machine Lock/Unlock}\label{fig:stateMachineLockUnlock}
\end{figure}
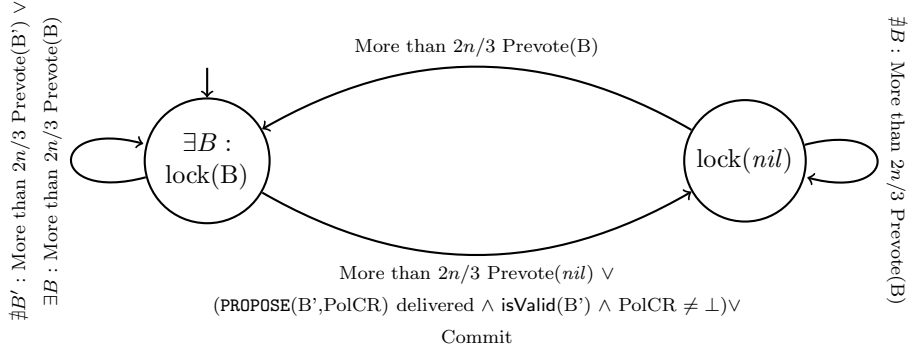
To preserve safety, when a validator delivers more than $2n/3$ prevotes for $B$ then it \vir{locks} on such block. Informally, it means that there are at least $n/3+1$ prevotes for $B$ from correct processes, then $B$ is a possible candidate for a decision so that validators try to stick on that. More formally, a validator has a \emph{Proof-of-LoCk} (PoLC) for a block $B$ (resp. for $\nil$) at a round $r$ for the height $H$ if it received
at least $2n/3+1$ prevotes for $B$ (resp. for $\nil$). In this case we say that a process is locked on such block. 
A \emph{PoLC-Round} (\PoLCR) is a round such that there was a PoLC for a block at round \PoLCR. In Figure \ref{fig:stateMachineLockUnlock} the state machine concerning the process of locking and unlocking on a block $B$ is shown. On the edges are reported the conditions on the delivered messages that have to be met in order to lock or unlock on a block. Intuitively, when a process delivers $2n/3$ of the same message $B$ of type prevote when in a precommit step, it locks on $B$. A process unlocks a block only if it delivers $2n/3$ of a value $B'$ or when it commits.
When a process is locked on a block $B$, it does not send any value different than $B$. This mechanism is necessary to ensure the safety of the protocol and to satisfy the Agreement property stated in Section \ref{sec:model}.
%
%


\textbf{Preamble.}
Note that our analysis of the original Tendermint protocol  \cite{Tendermint,buchman-thesis2016,know-manuscript2014} led to the conclusion that several modifications are needed in order to implement One-Shot Consensus problem. 
Full description of these bugs in the original Tendermint protocol are reported in Section \ref{sec:scenario}. 
In more details, with respect to the original Tendermint, our Tendermint One-shot Consensus algorithm (see Figure \ref{fig:code}) has the following modifications. We added line \ref{PV-end} in order to catch up  the communication delay during the synchronous periods. Moreover, we modified the line \ref{PV-unlock} in order to guarantee the agreement property of  One-Shot Consensus (defined in Section \ref{sec:model}). 
The correctness of  Tendermint One-shot Consensus algorithm needs an additional assumption stating that eventually a proposal is accepted by a majority of correct processes. This assumption, stated formally in Theorem \ref{t:consensus}, is necessary to guarantee the termination. 


\textbf{Variables and data structures.}
	$r$ and $\PoLCR_i$ are integers representing the current round and the PolCR.
	$\lockedBlock_i$ is the last block on which $p_i$ is locked, if it is equal to a block $B$, we say that $p_i$ is locked on $B$, otherwise it is equal to $\nil$, and we say that $p_i$ is not locked. When $\lockedBlock_i \neq \nil$ and switch the value to $\nil$, then $p_i$ unlocks.
	Last-Locked-Round ($\LLR_i$) is an integer representing the last round where $p_i$ locked on a block. $B$ is the block the process created.
	
	Each validator manages timeouts, $\TimeOutPropose$ and $\TimeOutPrevote$, concerning the propose and prevote phases respectively. Those timeouts are set to $\Delta_{\text{Propose}}$ and $\Delta_{\text{Prevote}}$ and started at the beginning of the respective step. Both are incremented if they expire before the validator moves to the next step. 
	
	Each validator manages three sets for the messages delivered. 
	In particular, the set
	$\proposalReceived^{H,r}_i$ contains the proposal that $p_i$ delivered for the round $r$ at height $H$.
	$\prevotesReceived^{H,r}_i$ is the set containing all the prevotes $p_i$ delivered for the round $r$ at height $H$.
	$\precommitsReceived^{H,r}_i$ is the set containing all the precommits $p_i$ delivered for the round $r$ at height $H$.
	\\
	\\
\textbf{Functions.}
	We denote by $\Block$ the set containing all blocks, and by $\MemPool$ the structure containing all the transactions.
	\begin{itemize}
		\item $\textsf{proposer}: V \times Height \times Round \to {V}$ is a deterministic function which gives the proposer out of the validators for a given round at a given height in a round robin fashion.
		\item $\textsf{createNewBlock}: 2^{\Pi} \times \MemPool \to \Block$ is an application-dependent function which creates a valid block (w.r.t. the application), where the subset of processes is a parameter of the One-Shot Consensus, and is the subset of processes that send a commit for the block at the previous height, called the signature of the previous block.
		\item $\textsf{is23Maj}: (Block \cup \nil) \times (\prevotesReceived\ \cup\ \precommitsReceived) \to \Bool$ is a predicate that checks if there is at least $2n/3+1$ of prevotes or precommits on the given block or $nil$ in the given set.
		\item $\textsf{isValid}: \Block \to \Bool$ is an application dependent predicate that is satisfied if the given block is valid. If there is a block $B$ such that $\textsf{isValid} (B) = \textsf{true}$, we say that $B$ is valid. We note that for any non-block, we set $\textsf{isValid}$ to false, (e.g. $\textsf{isValid}(\bot) = \textsf{false}$).
	\end{itemize}
	\leavevmode
	\\
\textbf{Detailed description of the algorithm.}
In Figure \ref{fig:code} we describe Tendermint One-Shot algorithm to solve the One-Shot Consensus (defined in Section \ref{sec:model}) for a given height $H$.

\noindent For each round $r$ at height $H$ the algorithm proceeds in 3 phases: 
\begin{enumerate}
	\item Propose step (lines \ref{Prop-begin} - \ref{Prop-end}):
		If $p_i$ is the proposer of the round and it is not locked on any block, then it creates a valid proposal and broadcasts it. Otherwise it broadcasts the block it is locked on.
		If $p_i$ is not the proposer then it waits for the proposal from the proposer.
		$p_i$ sets the timer to $TimeOutProposal$, if the timer expires before the delivery of the proposal then $p_i$ increases the time-out, otherwise it stores the proposal in $\proposalReceived^{H,r}_i$. In any case, $p_i$ goes to the Prevote step.
	
	\item Prevote step (lines \ref{PV-begin} - \ref{PV-end}):
		If $p_i$ delivers the proposal during the Propose step, then it checks the data on the proposal. 
		If $\lockedBlock_i \neq \nil$, and $p_i$ delivers a proposal  with a valid $\PoLCR$ then it unlocks. After that check, if $p_i$ is still locked on a block, then it prevotes on $\lockedBlock_i$;
		otherwise it checks if the block $B$ in the proposal is valid or not, if $B$ is valid, then it prevotes $B$, otherwise it prevotes on $\nil$.
		Then $p_i$ waits until $|\prevotesReceived^{H,r}_i| >2n/3$. If there is no PoLC for a block or for $\nil$ for the round $r$, then $p_i$ sets the timer to $\TimeOutPrevote$, waits for the timer's expiration and increases $\TimeOutPrevote$.
		In any case, $p_i$ goes to Precommit step.
		
	\item Precommit step (lines \ref{PC-01} - \ref{PC-wait}):
		$p_i$ checks if there was a PolC for a particular block or $\nil$ during the round (lines \ref{PC-01} and \ref{PC-03}). 
		There are three cases:
		(i) if there is a PoLC for a block $B$, then it locks on $B$, and precommits on $B$ (lines \ref{PC-01} - \ref{PC-02}); 
		(ii) if there is a PoLC for $\nil$, then it unlocks and precommits on $\nil$ (lines \ref{PC-03} - \ref{PC-04});
		(iii) otherwise, it precommits on $\nil$ (line \ref{PC-05});
		in any case, 
		$p_i$ waits until $|\precommitsReceived^{H,r}_i| > 2n/3$ or $({\sf is23Maj} (\nil,\prevotesReceived^{H,r}_i))$,
		and it goes to the next round.
\end{enumerate}

Whenever $p_i$ delivers a message, it broadcasts it (lines \ref{Prop-bc}, \ref{PV-bc} and \ref{PC-bc}). Moreover, during a round $r$, some conditions may be verified after a delivery of some messages and either (i) $p_i$ decides and terminates or (ii) $p_i$ goes to the round $r'$ (with $r'>r$). 
The conditions are:
\begin{itemize}
	\item For any round $r'$, if for a block $B$, $\textsf{is23Maj} (B,\precommitsReceived^{H,r'}_i) = \textsf{true}$ , then $p_i$ decides the block $B$ and terminates, or
	\item If $p_i$ is in the round $r$ at height $H$ and $|\prevotesReceived^{H,r'}_i| >2n/3$ where $r'>r$, then it goes to the Prevote step for the round $r'$, or
	\item If $p_i$ is in the round $r$ at height $H$ and $|\precommitsReceived^{H,r'}_i| >2n/3$ where $r'>r$, then it goes to the Precommit step for the round $r'$.
\end{itemize}

		\subsection{Correctness of Tendermint One-Shot Consensus}\label{ssec:proofConsensus}

In this section we prove the correctness of Tendermint One-Shot Consensus algorithm (Fig. \ref{fig:code}) for a height $H$ 
under the assumption that during the synchronous period there exists eventually a proposer such that its proposed value will be accepted by  at least $2n/3+1$ processes.
\setcounter{theorem}{0}
	\begin{lemma}[One-Shot Integrity]\label{l:integrity}
		In an eventual synchronous system, Tendermint One-Shot Consensus Algorithm verifies the following property:
		No correct process decides twice.
	\end{lemma}
	
	\begin{proofL}
		The proof follows by construction. A correct process decides when it returns (line \ref{return}).
		\renewcommand{\toto}{l:integrity}
	\end{proofL}
		
	\begin{lemma}[One-Shot Validity]\label{l:validity}
		In an eventual synchronous system, Tendermint One-Shot Consensus Algorithm verifies the following property:
		A decided value is valid, if it satisfies the predefined predicate denoted $\textsf{isValid}()$.
	\end{lemma}
	
	\begin{proofL}

		Let $p_i$ be a correct process, we assume that there exists a value $B$, such that $p_i$ decides $B$. 
		We show by construction that if $p_i$ decides on a value $B$, 
		then $B$ is valid. 
		
		If $p_i$ decides $B$, then $\textsf{is23Maj}(B,\precommitsReceived^{H,r}_i) = \textsf{true}$ (line \ref{return}), since the signature of the messages are unforgeable and $f < n/3$ by hypothesis, then more than $n/3$ of those precommits for round $r$ are from correct processes. This means that for each of those correct processes $p_j$, $\textsf{is23Maj}(B,\prevotesReceived^{H,r}_j) = \textsf{true}$ (lines \ref{PC-01} - \ref{PC-02}), and thus at least $n/3$ of those prevotes are from correct processes.
		
		Let $p_j$ be one of the correct processes which prevoted on $B$ during the round $r$. 
		$p_j$ prevotes on a value $B$ during a round $r$ in two cases: Case a, during $r$, $p_j$ is not locked on any value or Case b: $p_j$ is already locked on $B$ and does not checks its validity (lines \ref{PV-01} - \ref{PV-02}).
		\begin{itemize}
			\item Case a: If $p_j$ is not locked on any value than before prevoting it checks the validity of $B$ and prevotes on $B$ if $B$ is valid (line \ref{PV-valid01});
			\item Case b:  If $p_j$ was locked on $B$, it did not check the validity of $B$, it means that $p_j$ was locked on $B$ during the round $r$; which means that there was a round $r' < r$ such that $p_j$ had a PoLC for $B$ for the round $r'$ (lines \ref{PC-01} - \ref{PC-02}), by the same argument, there is a round which happened before $r'$ where $p_j$ was locked or $B$ is valid.
		
			Since a process locked during a round smaller than the current one, and that there exists a first round where all correct processes are not locked (line \ref{P-01}), there is a round $r'' < r'$ where  $p_j$ was not locked on $B$ but prevoted $B$, as in Case a, $p_j$ checks if $B$ is valid and then prevotes on $B$ if $B$ is valid (lines \ref{PV-valid01}).
		\end{itemize}
		A value prevoted by a correct process is thus valid. Therefore a decided value by a correct process is valid since more than $n/3$ correct processes prevote that value.
		\renewcommand{\toto}{l:validity}
	\end{proofL}

	\begin{lemma}\label{l:locklimit}
		In an eventual synchronous system, Tendermint One-Shot Consensus Algorithm verifies the following property:
		If $f+1$ correct processes locked on the same value $B$ during a round $r$ then no correct process can lock during round $r' > r$ on a value $B' \neq B$.
	\end{lemma}
	
	\begin{proofL}
		We assume that $f+1$ correct processes are locked on the same value $B$ during the round $r$, and we denote by $X^r$ the set of those processes.
		We first prove by induction that no process in $X^r$ will unlock or lock on a new value.
		Let let $p_i \in X^r$.
		\begin{itemize}
			\item \emph{Initialization}: round $r+1$.
				At the beginning of round $r+1$, all processes in $X^r$ are locked on $B$. Moreover, we have that $\LLR_i = r$, since $p_i$ locks on round $r$ (line \ref{PC-lock}).
				Let $p_j$ be the proposer for round $r+1$. If $\LLR_j = r$, it means that $p_j$ is also locked on $B$, since there cannot be a value $B' \neq B$ such that $\textsf{is23Maj}(B,\prevotesReceived^{H,r}_j) = \textsf{true}$, for that to happen, at least $n/3$ processes should prevote both $B$ and $B'$ during round $r$, which means that at least a correct process prevoted two times in the same round, which is not possible, since it is correct, and the protocol does not allow to vote two times in the same round (lines \ref{PV-begin} - \ref{PV-end}). Three cases can then happen:
				\begin{itemize}
					\item $p_j$ locked on a value $B_j$ during the round $\LLR_j \le r$. This  means that during the round $\LLR_j$ $\textsf{is23Maj}(B_j,\prevotesReceived^{H,\LLR_j}_j) = \textsf{true}$ (line \ref{PC-lock}). $p_j$ the proposer proposes a value $B_j$ along with $\LLR_j$ (lines \ref{Prop-begin} - \ref{Prop-bc1}). Since $\LLR_j \le \LLR_i = r$, $p_i$ does not unlock and prevotes $B$ for the round $r+1$, and so are  all the other processes in $X^r$ (lines \ref{PV-begin} - \ref{PV-01}). The only value that can have more than $2n/3$ prevotes is then $B$. So $p_i$ is still locked on $B$ at the end of $r+1$.
					\item If $p_j$ is not locked, the value it proposes cannot unlock processes in $X^r$ because $-1 = \LLR_j < r$, and they will prevote on $B$ (lines \ref{PV-begin} - \ref{PV-01}). The only value that can have more than $2n/3$ prevotes is then $B$. So $p_i$ is still locked on $B$ at the end of $r+1$.
					\item $p_j$ locked on a value $B_j$ during the round $\LLR_j > r$, $p_j$ the proposer proposes a value $B_j$ along with $\LLR_j$ (lines \ref{Prop-begin} - \ref{Prop-bc1}). Since $\LLR_j \ge r+1$, $p_i$ does not unlock and prevotes $B$ for the round $r+1$, and so are  all the other processes in $X^r$ (lines \ref{PV-begin} - \ref{PV-01}). The only value that can have more than $2n/3$ prevotes is then $B$. So $p_i$ is still locked on $B$ at the end of $r+1$.
				\end{itemize}
				At the end of round $r+1$, all processes in $X^r$ are still locked on $B$ and it may happen that other processes are locked on $B$ for round $r+1$ at the end of the round.
			
			\item \emph{Induction}: We assume that for a given $a > 0$, the processes in $X^r$ are still locked on $B$ at each round between $r$ and $r+a$. We now prove that the processes in $X^r$ will still be locked on $B$ at round $r+a+1$.
			
				Let $p_j$ be the proposer for round $r+a+1$. Since the $f+1$ processes in $X^r$ were locked on $B$ for all the rounds between $r$ and $r+a$, no new value can have more than $2n/3$ of prevotes during one of those rounds, so $\nexists B' \neq B : \textsf{is23Maj}(B',\prevotesReceived^{H,r_j}_j) = \textsf{true}$ where $r < r_j < r+a+1$. Moreover, if $p_j$ proposed the value $B$ along with a $\LLR > r$, since the processes in $X^r$ are already locked on $B$, they do not unlock and prevote $B$ (lines \ref{PV-begin} - \ref{PV-01}).
				The proof then follows as in the \emph{Initialization} case.
		\end{itemize}
		
		Therefore all processes in $X^r$ will stay locked on $B$ at each round after round $r$.
		Since $f+1$ processes will stay locked on the value $B$ on rounds $r' > r$, they will only prevote on $B$ (lines \ref{PV-begin} - \ref{PV-01}) for each new round.
		Let $B'$ be a value, we have that $\forall r' \ge r$ if $B': \textsf{is23Maj}(B',\prevotesReceived^{H,r'}_j) = \textsf{true}$ then $B' = B$.
		
		\renewcommand{\toto}{l:locklimit}
	\end{proofL}
	
	\begin{lemma}[One-Shot Agreement]\label{l:agreement}
		In an eventual synchronous system, Tendermint One-Shot Consensus Algorithm verifies the following property:
		If there is a correct process that decides a value $B$, then eventually all the correct processes decide $B$.
	\end{lemma}
	
	\begin{proofL}
		Let $p_i$ be a correct process. Without loss of generality, we assume that $p_i$ is the first correct process to decide, and it decides $B$ at round $r$.
		If $p_i$ decides $B$, then $\textsf{is23Maj}(B,\precommitsReceived^{H,r}_i) = \textsf{true}$ (line \ref{return}), since the signature of the messages are unforgeable by hypothesis and $f < n/3$, then $p_i$ delivers more than $n/3$ of those precommits for round $r$ from correct processes, and those correct process are locked on $B$ at round $r$ (line \ref{PC-lock}).
		$p_i$ broadcasts all the precommits it delivers (line \ref{PC-bc}), so eventually
		all correct processes will deliver those precommits, because of the best effort broadcast guarantees.
		
		We now show that before delivering the precommits from $p_i$, the other correct processes cannot decide a different value than $B$.
		$f < n/3$ by hypothesis, so we have that at least $f+1$ correct processes are locked on $B$ for the round $r$.
		By Lemma \ref{l:locklimit} no correct process can lock on a value different than $B$. Let $B' \neq B$, since correct processes lock only when they precommit (lines \ref{PC-01} - \ref{PC-02}), no correct process will precommit on $B'$ for a round bigger than $r$, so
		$\textsf{is23Maj}(B',\precommitsReceived^{H,r'}_i) = \textsf{false}$ for all $r' \ge r$ since no correct process will precommit on $B'$. No correct process cannot decide a value $B' \neq B$ (line \ref{return}) once $p_i$ decided.
		Eventually, all the correct processes will deliver the $2n/3$ signed precommits $p_i$ delivered and broadcasted, thanks to the best effort broadcast guarantees and then will decide $B$.
		\renewcommand{\toto}{l:agreement} 
	\end{proofL}
	
	\begin{lemma} \label{l:termination_lemma}
		In an eventual synchronous system, and under the assumption that during the synchronous period eventually there is a correct proposer $p_k$ such that $|\{p_j: \LLR_k \le \LLR_j \text{ and $p_j$ is correct}\}| < n/3 - f$, Tendermint One-Shot Consensus Algorithm verifies the following property:
		Eventually a correct process decides.
		
	\end{lemma}
	\begin{proofL}
		Let $r$ be the round where the communication becomes synchronous and when all the messages broadcasted by correct processes are delivered by the correct processes within their respective step.
		The round $r$ exists, since the system is eventually synchronous and correct processes increase their time-outs when they did not deliver enough messages (lines \ref{Prop-01} - \ref{Prop-02}, \ref{PV-timer} - \ref{PV-end} and \ref{PC-wait}).
		If a correct process decides before $r$, that ends the proof. Otherwise no correct process decided yet. Let $p_i$ be the proposer for the round $r$.
		We assume that $p_i$ is correct. Let $B$ be the value such that $p_i$ proposes $(B, \LLR_i)$, we have three cases:
		\begin{itemize}
			\item {Case 1:} No correct process is locked on a value before $r$. $\forall p_j \in \Pi$ such that $p_j$ is correct, $\LLR_j = -1$.
			
				Correct processes delivered the proposal $(B, \LLR_i)$ before  the Prevote step (lines \ref{Prop-wait}, \ref{del-Prop01} - \ref{Prop-bc}).
				Since the proposal is valid, then all correct processes will prevote on that value (line \ref{PV-valid01}), and they deliver the others' prevotes and broadcast them before entering the Precommit step (lines \ref{PV-wait} - \ref{PV-end} and \ref{PV-bc}).
				Then for all correct process $p_j$, we have $\textsf{is23Maj}(B,\precommitsReceived^{H,r}_j) = \textsf{true}$. The correct processes will lock on $B$, precommit on $B$ (lines \ref{PC-01} - \ref{PC-02}) and will broadcast all precommits delivered (line \ref{PC-bc}).
				Eventually a correct process $p_j$ will have $\textsf{is23Maj}(B,\precommitsReceived^{H,r}_j) = \textsf{true}$ then $p_j$ will decide (line \ref{return}).
				
			\item {Case 2:} Some correct processes are locked and if $p_j$ is a correct process, $\LLR_j < \LLR_i$.
			
				Since $\LLR_j < \LLR_i$ for all correct processes $p_j$, then the correct processes that are locked will unlock (line \ref{PV-unlock}) and the proof follows as in the Case 1. 
				
			\item {Case 3:} Some correct processes are locked on a value, and there exist a correct process $p_j$ such that $\LLR_i \le \LLR_j$.
				\begin{itemize}
					\item {(i)} If $|\{p_j: \LLR_i \le \LLR_j \text{ and $p_j$ is correct}\}| < n/3 - f$ (which means that even without the correct processes that are locked in a higher round than the proposer $p_i$, there are more than $2n/3$ other correct processes unlock or locked in a smaller round than $\LLR_i$), then as in the case 2, a correct process will decide.
					\item {(ii)} If $|\{p_j: \LLR_i \le \LLR_j \text{ and $p_j$ is correct}\}| \ge n/3 - f$, then during the round $r$, $\nexists B': \textsf{is23Maj}(B',\precommitsReceived^{H,r}_i) = \textsf{true}$, in fact correct processes only precommit once in a round (lines \ref{PC-01} - \ref{PC-wait}). 
Eventually, thanks to the additional assumption, there exists a round $r_1$ where the proposer $p_k$  is correct and at round $r_1$, $|\{p_j: \LLR_k \le \LLR_j \text{ and $p_j$ is correct}\}| < n/3 - f$. The proof then follows as case (3.i).
				\end{itemize}
		\end{itemize}
		If $p_i$ is Byzantine and more than $n/3$ correct processes delivered the same message during the proposal step, and the proposal is valid, the situation is like $p_i$ was correct.
		Otherwise, there are not enough correct processes that delivered the proposal, or if the proposal is not valid, then there will be less than $n/3$ processes that will prevote that value. No value will be committed. Since the proposer is selected in a round robin fashion, a correct process will eventually be the proposer, and a correct process will decide.
		\renewcommand{\toto}{l:termination_lemma}
	\end{proofL}
	
	\begin{lemma}[One-Shot Termination]\label{l:termination}
		In an eventual synchronous system, and under the assumption that during the synchronous period eventually there is a correct proposer $p_k$ such that $|\{p_j: \LLR_k \le \LLR_j \text{ and $p_j$ is correct}\}| < n/3 - f$, Tendermint One-Shot Consensus Algorithm verifies the following property:
		Every correct process eventually decides some value.
	\end{lemma}
	
	\begin{proofL}
		By construction, if a correct process does not deliver a proposal during the proposal step or enough prevotes during the Prevote step, then that process increases its time-outs (lines \ref{Prop-01} - \ref{Prop-02} and \ref{PV-timer} - \ref{PV-end}), so eventually, during the synchrony period of the system, all the correct processes will deliver the proposal and the prevotes from correct processes respectively during the Propose and the Prevote step. 
		By Lemma \ref{l:termination_lemma}, a correct process decides a value, and then by the Lemma \ref{l:agreement}, every correct process eventually decides.
		\renewcommand{\toto}{l:termination}
	\end{proofL}
	\begin{theorem}\label{t:consensus}
		In an eventual synchronous system, and under the assumption that during the synchronous periods eventually there is a correct proposer $p_k$ such that $|\{p_j: \LLR_k \le \LLR_j \text{ and $p_j$ is correct}\}| < n/3 - f$:
		Tendermint One-Shot Algorithm implements the One-Shot Consensus.
	\end{theorem}
	
	\begin{proofT}
		The proof follows directly from Lemmas \ref{l:integrity},  \ref{l:validity}, \ref{l:agreement} and \ref{l:termination}.
		\renewcommand{\toto}{t:consensus}
	\end{proofT}
		\subsection{Tendermint Repeated Consensus algorithm}\label{ssec:algoRepeatedConsensus}

\begin{figure*}[t!]
	\centering
	\fbox{
		\begin{minipage}{0.4\textwidth}
			\scriptsize
			\resetline
			\begin{tabbing}
				aaaA\=aA\=aA\=aA\=aA\=aaaA\kill
				{\bf Function} {\sf repeatedConsensus}$(\Pi)$; \%Repeated Consensus for the set $\Pi$ of processes\%\\ \\
				
				{\bf Init}:\\
					\line{} \> $H \leftarrow 1$ \%\Height\%; $B \leftarrow \bot;$ $V \leftarrow \bot$ \%Set of validators\%;\\
					\line{} \> $\commitsReceived^{H}_i \leftarrow \emptyset$; $\toReward^{H}_i \leftarrow \emptyset$; $\TimeOutCommit \leftarrow \Delta_{\text{Commit}};$\\
				~---------------------------------------------------------------------------------------------------------------\\
				{\bf while} ${\sf (true)}$ {\bf do}\\
					\line{repC:init} \> $B \leftarrow \bot;$\\
					\line{} \> $V \leftarrow validatorSet(H);$ \%Application and blockchain dependant\% \\
					\line{repC:val} \>  {\bf if} $(p_i \in V)$ {\bf then}\\
						\line{repC:cons} \>\> $B \leftarrow \textsf{consensus}(H,V,\toReward^{H-1}_i);$ \%Consensus function for the height $H$\%\\
						\line{repC:bc} \>\> {\bf trigger} {\sf broadcast} $\langle \texttt{COMMIT}, (B,H)_i \rangle$;\\
					\line{} \> \  {\bf else} \\
						\line{repC:wait} \>\> \textbf{wait until} $(\exists B': |\textsf{atLeastOneThird}(B',\commitsReceived^{H}_i)|)$;\\
						\line{} \>\> $B \leftarrow B'$;\\
					\line{} \> {\bf endif}\\
					\line{} \> $\textsf{set } timerCommit \textsf{ to } \TimeOutCommit;$\\
					\line{Commit-wait} \> $\textbf{wait until} (timerCommit \text{ expired});$\\
					\line{repC:output} \> $\textbf{trigger }\textsf{decide} (B);$\\
					\line{} \> $H \leftarrow H + 1;$\\
				{\bf endwhile} \\
				~---------------------------------------------------------------------------------------------------------------\\
				 {\bf upon event} {\sf delivery} $\langle \texttt{COMMIT}, ( B',H')_j \rangle$:\\
				 	\line{ComDel-01} \>  {\bf if} $((( B',H')_j \notin \commitsReceived^{H'}_i) \land (p_j \in validatorSet(H')))$ {\bf then}\\
						\line{} \>\> $\commitsReceived^{H'}_i \leftarrow \commitsReceived^{H'}_i \cup ( B',H')_j$;\\
						\line{Commit:reward} \>\> $\toReward^{H'}_i \leftarrow \toReward^{H'}_i \cup p_j$;\\
						\line{Commit:bc} \>\> {\bf trigger} {\sf broadcast} $\langle \texttt{COMMIT}, ( B',H')_j \rangle$;\\
					\line{ComDel-02} \> {\bf endif}\\
			\end{tabbing}
			\normalsize
		\end{minipage}%
	}
	\caption{Tendermint Repeated Consensus algorithm at correct process $p_i$.}
	\label{fig:repConsensus}
\end{figure*}

For a given height, the set $V$ of validators does not change. Note that  each height corresponds to a block. Therefore, in the following we refer this set as the set of validators for a block.
\\
\\
\textbf{Data structures.}
The integer $H$ is the height where is called a One-Shot Consensus instance.
$V$ is the current set of validators.
$B$ is the block to be appended.
$\commitsReceived^{H}_i$ is the set containing all the commits $p_i$ delivered for the height $H$. $\toReward^{H}_i$ is the set containing the validators from which $p_i$ delivered commits for the height $H$.
$\TimeOutCommit$ represents the time a validator has for collecting commits after an instance of consensus. $\TimeOutCommit$ is set to $\Delta_{\text{Commit}}$.
\\
\\
\textbf{Functions.}
	\begin{itemize}
		\item $\textsf{validatorSet}: {\Pi \times} \Height \to 2^\Pi$ is an application dependent and  deterministic selection function which gives the set of validators for a given height w.r.t the blockchain history. We have $\forall H \in \Height, |validatorSet(H)| = n$.
		\item $\textsf{consensus}:\Height \times 2^\Pi \times \commitsReceived \to \Block$ is the One-Shot Consensus instance presented in \ref{ssec:algoConsensus}.				
		\item $\textsf{createNewBlock}: 2^{\Pi} \times \MemPool \to \Block$ is the application-dependent function that creates a valid block (w.r.t. the application) from the One-Shot Consensus. 
		\item $\textsf{atLeastOneThird}: \Block \times \commitsReceived \to \Bool$ is a predicate which checks if there is at least $n/3$ of commits of the given block in the given set.
		\item $\textsf{isValid}: \Block \to \Bool$ is the same predicate as in the One-shot Consensus, which checks if a block is valid or not.
	\end{itemize}
	\leavevmode
	\\
\textbf{Detailed description of the algorithm.}
	In Fig. \ref{fig:repConsensus} we describe the algorithm to solve the Repeated Consensus as defined in Section \ref{sec:model}. The algorithm proceeds as follows: 
	\begin{itemize}
		\item $p_i$ computes the set of validators for the current height;
		\item If $p_i$ is a validator, then it calls the consensus function solving the consensus for the current height, then broadcasts the decision, and sets $B$ to that decision;
		\item Otherwise, if $p_i$ is not a validator, it waits for at least $n/3$ commits from the same block and sets $B$ to that block;
		\item In any case, it sets the timer to $\TimeOutCommit$ for receiving more commits and lets it expire. Then $p_i$ decides $B$ and goes to the next height.
	\end{itemize}
	Whenever $p_i$ delivers a commit, it broadcasts it (lines \ref{ComDel-01} - \ref{ComDel-02}).
	Note that the reward for the height $H$ is given during the height $H+1$, and to a subset of validators who committed the block for $H$ (line \ref{repC:cons}).

		\subsection{Correctness of Tendermint Repeated Consensus}\label{ssec:proofRepeatedConsensus}

In this section we prove the correctness of Tendermint Repeated Consensus algorithm in Figure \ref{fig:repConsensus}.
We now show that Tendermint Repeated Algorithm (Fig. \ref{fig:repConsensus}) implements the Repeated Consensus. 

\begin{lemma}[Repeated Termination]\label{l:terminationRep}
	In an eventual synchronous system, and under the additional assumption that during the synchronous period eventually there is a correct proposer $p_k$ such that $|\{p_j: \LLR_k \le \LLR_j \text{ and $p_j$ is correct}\}| < n/3 - f$, Tendermint Repeated Consensus Algorithm verifies the following property:
	Every correct process has an infinite output. 
\end{lemma}

\begin{proofL}
	By contradiction, let $p_i$ be a correct process, and we assume that $p_i$ has a finite output.
	Two scenarios are possible, either $p_i$ cannot go to a new height, or from a certain height $H$ it outputs only $\bot$.
	\begin{itemize}
		\item If $p_i$ cannot progress, one of the following cases is satisfied:
			\begin{itemize}
				\item The function $\textsf{consensus}()$ does not terminate (line \ref{repC:cons}), which is a contradiction due to Lemma \ref{l:termination}; or
				\item $p_i$ waits an infinite time for receiving enough commits (line \ref{repC:wait}), which cannot be the case because of the best effort broadcast guarantees and the eventual synchronous assumption, all the correct validators terminate the One-Shot Consensus and broadcast their commit. 
			\end{itemize}
		\item If $p_i$ decides at each height (line \ref{repC:output}), it means that from a certain height $H$,  $p_i$ only outputs $\bot$. That means that: 
			(i) either $p_i$ is a validator for $H$ and the function $\textsf{consensus}(H')$ is only returning $\bot$ for all $H' \ge H$ (lines \ref{repC:val} and \ref{repC:cons}), or
			(ii) $p_i$ is not a validator for $H$ but delivered at least $n/3$ commits for $\bot$ (lines \ref{repC:wait} and \ref{Commit:bc}).
			\begin{itemize}
				\item (i): Since $\textsf{consensus}()$ returns the value $\bot$, that means by Lemma \ref{l:validity} that $\textsf{isValid}(\bot) = \textsf{true}$, which is a contradiction with the definition of the function $\textsf{isValid}()$.
				\item (ii): Since only the validators commit, and each of them broadcasts its commit (lines \ref{repC:val} - \ref{repC:bc}), and  because $f < n/3$, it means that $p_i$ delivered a commit from at least one correct validator (process). By Lemma \ref{l:validity}, correct processes only decide/commit on valid value, and $\bot$ is not valid, which is a contradiction.
			\end{itemize}
	\end{itemize}
	We conclude that if $p_i$ is a correct process, then it has an infinite output.
	\renewcommand{\toto}{l:terminationRep}
\end{proofL}

\begin{lemma}[Repeated Agreement]\label{l:agreementRep}
	In an eventual synchronous system, Tendermint Repeated Consensus Algorithm verifies the following property:
	If the $i^{th}$ value of the output of a correct process is $B$, then $B$ is the $i^{th}$ value of the output of any other correct process.
\end{lemma}

\begin{proofL}
	We prove this lemma by construction. Let $p_j$ and $p_k$ be two correct processes. 
	Two cases are possible:
	\begin{itemize}
		\item $p_j$ and $p_k$ are validators for the height $i$, so they call the function $\textsf{consensus}()$ (lines \ref{repC:val} and \ref{repC:cons}). By Lemma \ref{l:agreement} $p_j$ and $p_k$ decide the same value and then output that same value (line \ref{repC:output}).
		\item At least one of $p_j$ and $p_k$ is not a validator for the height $i$. Without loss of generality, we assume that $p_j$ is not a validator for the height $i$. Since all the correct validators commit the same value, let say $B$, thanks to Lemma \ref{l:agreement}, and since they broadcast their commit (line \ref{repC:bc}), eventually there will be more than $2n/3$ of commits for $B$. So no other value $B' \neq B$ can be present at least $n/3$ times in the set $commitReceived^{H}_i$. So $p_j$ outputs the same value $B$ as all the correct validators (line \ref{repC:wait}). If $p_k$ is a validator, that ends the proof. If $p_k$ is not a validator, then by the same argument as for $p_j$, $p_k$ outputs the same value $B$. Hence $p_j$ and $p_k$ both output the same value $B$.
			
	\end{itemize}
	\renewcommand{\toto}{l:agreementRep}
\end{proofL}

\begin{lemma}[Repeated Validity]\label{l:validityRep}
	In an eventual synchronous system, Tendermint Repeated Consensus Algorithm verifies the following property:
	Each value in the output of any correct process is valid, it satisfies the predefined predicate denoted $\textsf{isValid}()$.
\end{lemma}

\begin{proofL}
	\renewcommand{\toto}{l:validityRep}
	We prove this lemma by construction. Let $p_i$ be a correct process, and we assume that the $H^{th}$ value of the output of $p_i$ is $B$.
	If $p_i$ decides a value (line \ref{repC:output}), then that value has been set during the execution and for that height (line \ref{repC:init}). 
	\begin{itemize}
		\item If $p_i$ is a validator for the height $H$, then $B$ is the value returned by the function $\textsf{consensus}()$, by the Lemma \ref{l:validity} we have that $\textsf{isValid}(B) = \textsf{true}$.
		\item If $p_j$ is not a validator for the height $H$, it means that it delivered more than $n/3$  signed commits from the validators for the value $B$ (lines \ref{repC:val} - \ref{repC:bc} and \ref{ComDel-01} - \ref{ComDel-02}), hence at least one correct validator committed $B$, and by Lemma \ref{l:validity} we  have that $\textsf{isValid}(B) = \textsf{true}$.
	\end{itemize}
	So each value that a correct process outputs satisfies the predicate $\textsf{isValid}()$.
\end{proofL}

\begin{theorem}\label{t:repConsensus}
	In an eventual synchronous system, Tendermint Repeated Consensus algorithm implements the Repeated Consensus.
	In an eventual synchronous system, and under the additional assumption that during the synchronous period eventually there is a correct proposer $p_k$ such that $|\{p_j: \LLR_k \le \LLR_j \text{ and $p_j$ is correct}\}| < n/3 - f$, Tendermint Repeated Consensus algorithm implements the repeated consensus.
\end{theorem}

\begin{proofT}
	The proof follows directly from Lemmas \ref{l:terminationRep}, \ref{l:agreementRep} and \ref{l:validityRep}.
	we showed that Tendermint protocol satisfies respectively the Termination property, the Agreement property and the Validity property.
	\renewcommand{\toto}{t:repConsensus}
\end{proofT}
	\section{Bugs in the original Tendermint}\label{sec:scenario}

\newcommand \assumption{\ensuremath{\mathcal{T}}}

\subsection{Addition of line \ref{PV-end} on Fig. \ref{fig:code}}
\label{ref:scenario}
	The line \ref{PV-end} allows the correct process $p_i$ to increase its time-out to catch up the communication delay in the network.
	If the correct processes never increase their $\TimeOutPrevote$, even when the system becomes synchronous, the correct process may never deliver enough prevotes at time. Thus it can never precommit.
	To decide, a correct process needs more than $2n/3$ precommits for the same value for the same round (and so more that $n/3$ precommits from correct processes) to decide (line \ref{return}), no correct process ever decides, which does not satisfies the One-Shot Termination property.
	When a process increases its \TimeOutPrevote whenever it does not deliver enough messages, it will eventually catch-up the delay, and during the synchronous period, there will be a time from when it will deliver prevotes from all correct processes.

\subsection{Modification of line \ref{PV-unlock} on Fig. \ref{fig:code}}
	Originally, the line \ref{PV-unlock} was \[{\bf if } (\exists B' : ({\sf is23Maj} (B',\prevotesReceived^{H,\PoLCR_i}_i)) \textbf{ then }\lockedBlock_i \leftarrow \nil; {\bf endif}\]
	In that version, it is possible that if a process $p_i$ is locked on a value $B$ during a round $r$, a process locked on the same value $B$ during a round $r' > r$ makes $p_i$ unlock, but does not ensure that $p_i$ locks again.
	That is a problem since it causes a violation of the Agreement property.
	
	In the following, we exhibit a problematic scenario.
	Assume that there are $4$ processes in the network. $3$ correct processes $p_1, p_2, p_3$ and a Byzantine process $p_4$.
	\begin{enumerate}
		\item Round 1: $p_1$ is the proposer and proposes $B$. All process deliver the proposal and prevote on $B$ from round 1.
		
			$p_1$ and $p_2$ deliver all prevotes for $B$, and then lock and precommit on $B$.
			
			$p_1$ delivers the precommit of $p_1, p_2$ and from $p_4$ for $B$and then it decides $B$. Neither $p_2$ and $p_3$ delivers enough precommit to decide.
			
			The state is: $p_1$ decides $B$ and left. $p_2$ is locked on $(B,1)$, $p_3$ is not locked, $p_4$ is Byzantine so we do not say anything about its state.
			
		\item Round 2: $p_1$ exit, and do not take part any more. $p_2$ and $p_3$ do not deliver the precommit for $B$.
		
			$p_2$ is the proposer. Since $p_2$ is locked on $B$, it proposes $B$ along with $1$ where it locked. $p_2$ and $p_3$ deliver the proposal.
			$p_2$ does not unlock since it locked at round $1$ and prevote on $B$.
			$p_3$ is not locked, but since it delivered the proposal, it prevotes on $B$.
			$p_4$ sends a prevote on $B$ only to $p_3$ such that $p_3$ delivers all the prevotes during this step but not $p_2$, so during the precommit step, $p_3$ locked on $B$ but for the round $2$.
			
			The state is: $p_1$ already decided. $p_2$ is locked on $(B,1)$, $p_3$ is locked on $(B,2)$, $p_4$ is Byzantine so we do not say anything about its state.
			
		\item Round 3: $p_2$ and $p_3$ do not deliver the precommit for $B$ from round 1.
		
			$p_3$ is the proposer. Since $p_3$ is locked on $B$, it proposes $B$ along with $2$ where it locked. $p_2$ and $p_3$ deliver the proposal.
			$p_2$ unlocks since it receives $2$ but was locked at $1$ and prevote the proposal $B$.
			$p_3$ does not unlock since it locked exactly at round $1$ and prevote on $B$.
			$p_4$ sends a prevote on $B$ such that $p_3$ delivers all the prevotes during its step but not $p_2$, so during the precommit step, $p_3$ locked on $B$ but for the round $2$.
			
			The state is: $p_1$ already decided $B$. $p_2$ is not locked, $p_3$ is locked on $(B,2)$, $p_4$ is Byzantine so we do not say anything about its state.	
			
		\item Round 4: $p_2$ and $p_3$ do not deliver the precommit for $B$ from round 1.
		
			$p_4$ is the proposer and proposes $(B,3)$.
			$p_2$ is not locked, but since it delivered the proposal, it prevotes on $B$.
			$p_3$ unlocks since it receives $3$ but was locked at $2$ and prevote the proposal $B$.
			$p_4$ does nothing. Neither $p_2$ nor $p_3$ delivers enough prevotes to lock.
			
			The state is: $p_1$ already decided $B$. $p_2$ is not locked, $p_3$ is not locked, $p_4$ is Byzantine so we do not say anything about its state.
		
		\item Round 5: $p_2$ and $p_3$ do not deliver the precommit for $B$ from round 1.
		
			$p_1$ is the proposer but since it left, there is no proposal.
			$p_2$ and $p_3$ are not locked and did not deliver a proposal so they prevote on $\nil$.
			$p_4$ does nothing. Neither $p_2$ nor $p_3$ delivers enough prevotes to lock.
			
			The state is: $p_1$ already decided $B$. $p_2$ is not locked, $p_3$ is not locked, $p_4$ is Byzantine so we do not say anything about its state.
			
		\item Round 4: $p_2$ and $p_3$ do not deliver the precommit for $B$ from round 1.
		
			$p_2$ is the proposer. Since it is not locked, it proposes a new value $B'$.
			$p_2$ and $p_3$ deliver the proposal before their respective prevote step.
			$p_2$ and $p_3$ are not locked and did deliver the proposal $B'$, so they prevote on $B'$.
			$p_4$ sends prevote $B'$ to both $p_2$ and $p_3$.
			
			$p_2$ and $p_3$ deliver the $3$ prevotes before entering the precommit step, and hence locked on $B'$ and precommit on $B$, and $p_4$ also send precommit to $p_2$.
			
			$p_2$ delivers precommit from $p_2, p_3$ and $p_4$ and thus it decides $B'$.			
						
			The state is: $p_1$ already decided $B$. $p_2$ decides $B'$, $p_3$ is locked on $B'$, $p_4$ is Byzantine so we do not say anything about its state.		
	\end{enumerate}
	At the end of the round 4, $p_1$ and $p_2$ decide on two different values, which does not satisfy the One-Shot Agreement property.
	
\subsection{Counter-example for the One-Shot Termination without the additional assumption}
\label{sec:counterexampletermination}
	
	We recall the additional assumption : eventually there is a correct proposer $p_k$ such that $|\{p_j: \LLR_k \le \LLR_j \text{ and $p_j$ is correct}\}| < n/3 - f$.

	In \cite{cachin2017blockchains} the authors advocate without providing any evidence that there is a livelock problem in Tendermint description proposed in the Buchman's manuscript \cite{buchman-thesis2016}. Hereafter, we exhibit this evidence. We thank anonymous reviewers of  the preliminary version of this work appeared as technical report \cite{ADPT18} to point us the scenario below. 

	
	We consider a system of 4 processes, $p_1$ to $p_4$, where $p_4$ is a Byzantine process. The round number 1 has $p_1$ as a proposer, and we assume that it happens before the system is synchronous (before GST in DLS terminology \cite{DLS88}), and only $p_1$ locks value $v_1$ in this round ($\lockedBlock = v_1,LLR = 1$ at $p_1$, and for other processes equal to initial values. $\PoLCR = \bot$ for $p_1$, $p_2$, and $p_3$ ($p_4$ is faulty so we don't talk about it's state).
	
	Starting with round number 2, all rounds happen during synchronous period (after GST), so communication between correct processes is reliable and timely, i.e., all correct processes receive messages from all correct processes. Note that during synchronous period we don't have the same guarantee on messages sent by Byzantine processes, i.e., a Byzantine process can send a message only to a subset of correct processes on time so it is delivered in the current round, and although a message is eventually delivered by all correct processes, it might not be delivered by all correct processes in the round $r$ in which it is sent.
 
	\begin{enumerate}
		\item In round 2, $p_2$ proposes $(v_2, \bot)$, where $B=v_2$, and $\PoLCR=\bot$ as $p_2$ hasn't locked any value.
			$p_1$ rejects this proposal and Prevote $v_1$ as it has $v_1$ locked in round 1 (condition at line \ref{PV-begin} evaluates to false as upon receipt of Proposal all correct processes set $\PoLCR$ to $\bot$). $p_2$ and $p_3$ accept the proposal and prevote $v_2$, but as $p_4$ stay silent, no process locks a value in round 2. So at the end of round 2, we have the following state: $\lockedBlock = v_1, \LLR = 1$ at $p_1$, and for other processes equal to initial values ($\lockedBlock = \nil$ and $\LLR = \bot$) $\PoLCR = \bot$ at $p_1$, $p_2$, and $p_3$ ($p_4$ is faulty so we don't talk about it's state).

	\item In round 3, $p_3$ proposes $(v_3, \bot)$ as it hasn't locked any value.
		$p_1$ rejects this proposal and prevote $v_1$ as it has $v_1$ locked in round 1 (condition at line \ref{PV-begin} evaluates to false as upon receipt of Proposal all correct processes set $\PoLCR$ to $\bot$). $p_2$ and $p_3$ accept the proposal and prevote $v_3$, but $p_4$ sends Prevote message for $v_3$ only to $p_3$. Furthermore, $p_4$ sends Prevote $v_3$ message to $p_3$ just before $timerPrevote$ expires at $p_3$, and after $timerPrevote$ expired at $p_1$ and $p_2$, and they moved to round 4. So although Prevote messages that are received by $p_3$ are propagated to other processes, they will be received by $p_1$ and $p_2$ after they moved to round 4. So in the round 3 only $p_3$ locks $v_3$. At the end of the round 3, $\lockedBlock = v_1, \LLR = 1$ at $p_1$, $\lockedBlock = v_3, \LLR = 3$ at $p_3$, $\lockedBlock = \nil, \LLR = \bot$ at $p_2$, and $\PoLCR = \bot$ for $p_1$, $p_2$, and $p_3$ ($p_4$ is faulty so we don't talk about it's state).

 	\item In round 4, $p_4$ is a proposer and as it is Byzantine process we assume
 it stays silent, so nothing change.

 	\item In round 5, $p_1$ proposes $(v_1, 1)$. 
		$p_3$ rejects this proposal and prevote $v_3$ as it has $v_3$ locked in round 3, therefore the condition at line \ref{PV-begin} evaluates to false as $\LLR = 3$ and $\PoLCR = 1$ at $p_3$. $p_1$ and $p_2$ accept the proposal, and prevote $v_1$. $p_4$ sends Prevote message for $v_1$ only to $p_1$ just before $timerPrevote$ expires at $p_1$ and after $timerPrevote$ expires at $p_2$ and $p_3$. So $p_2$ and $p_3$ receives Prevote from $p_4$ from round 5 only after they moved to round 6. So in the round 5 only $p_1$ locks $v_1$. At the end of the round 5, $\lockedBlock = v_1, \LLR = 5$ at $p_1$, $\lockedBlock = v_3, \LLR = 3$ at $p_3$, $\lockedBlock = \nil, \LLR = \bot$ at $p_2$, and $\PoLCR = 1$ for $p_1$, $p_2$, and $p_3$ ($p_4$ is faulty so we don't talk about it's state).

 	\item In round 6, $p_2$ proposes $(v_2, 1)$ as it hasn't locked any value.
		$p_1$ rejects this proposal and prevote $v_1$ as it has $v_1$ locked in round 5. $p_3$ rejects the proposal and prevote $v_3$ as it has locked $v_3$ in round 3, and $p_4$ stays silent, so no process lock a value in round 2. At the end of the round 6, $\lockedBlock = v_1, \LLR = 5$ at $p_1$, $\lockedBlock = v_3, \LLR = 3$ at $p_3$, $\lockedBlock = \nil, \LLR = \bot$ at $p_2$, and $\PoLCR = 1$ for $p_1$, $p_2$, and $p_3$ ($p_4$ is faulty so we don't talk about it's state).

 	\item In round 7, $p_3$ proposes $(v_3, 3)$.
		$p_1$ rejects this proposal and prevote $v_1$ as it has $v_1$ locked in round 5. $p_2$ and $p_3$ accept the proposal and prevote $v_3$, and $p_4$ sends Prevote message for $v_3$ only to $p_3$, just before $timerPrevote$ expires at $p_3$ and after $timerPrevote$ expires at $p_1$ and $p_2$, and after they moved to round 8. So similar as above in the round 7 only $p_3$ locks $v_3$. At the end of the round 7, $\lockedBlock = v_1, \LLR = 5$ at $p_1$, $\lockedBlock = v_3, \LLR = 7$ at $p_3$, $\lockedBlock = \nil, \LLR = \bot$ at $p_2$, and $\PoLCR = 3$ for $p_1$, $p_2$, and $p_3$ ($p_4$ is faulty so we don't talk about it's state).
	\end{enumerate}
	
 This scenario repeats forever, so the algorithm never terminates, which violates the One-Shot Termination property. This scenario cannot repeat forever with the additional assumption that during the synchronous period eventually there is a correct proposer $p_k$ such that $|\{p_j: \LLR_k \le \LLR_j \text{ and } p_j \text{ is correct}\}| < n/3 - f$.
 
 Let us call that assumption assumption $\assumption$. 
\paragraph{Assumption $\assumption$vs. Symmetric Byzantine}: Symmetric Byzantine processes \cite{KA94} are processes that behave arbitrarily, but their behaviour is perceived the same by all correct processes. We note that in our case; the assumption of Symmetric Byzantine is stronger than the assumption $\assumption$.
In fact the Symmetric Byzantine assumption restrict the Byzantine behaviour during the whole execution of the algorithm, whereas the assumption $\assumption$ requires that eventually, during an interval of one round of the execution after the synchronous period,  Byzantine processes behaviour do not impact at least $2n/3$ correct processes.

	\section{Tendermint Fairness}\label{sec:fairness}

Recently Pass and Shi defined the fairness of a Proof-of-Work based blockchain protocol for a system of $n$ processes as follows (please refer to \cite{PS17} for the formal definition): 
	\emph{A blockchain protocol is fair if honest process that wield $\phi$ fraction of the computational resources will reap at least $\phi$ fraction of the blocks in any sufficiently long window of the chain}, where the computational resources represent the merit of the process.
	We note that in their model, a block in the blockchain was created by only one process, and that process gets a reward for the created block. We extend the definition of \cite{PS17} for a system with an infinite number of processes, and where each block is produced  by a subset of processes. This is the case of  Tendermint for example where for each block there is a subset of processes called the \emph{validators} that produce that block. The \emph{correct validators} for a block (those that followed the protocol and participated in the agreement process) are the processes that have to be rewarded for that block. Informally, we say that a blockchain protocol is fair if any \emph{correct process} (a process that followed the protocol) that wield $\phi$ fraction of the total merit in the system will get more or less $\phi$ fraction of the total reward that is given in the system. In order to study the fairness of a protocol in a consensus-based blockchain such as Tendermint,  Redbelly, SBFT or Hyperledger Fabric, we split the protocol in two mechanisms: (i) the \emph{selection mechanism} which selects for each new height the \emph{validators} (the processes that will run the consensus instance) for that height taking into account the merit of each  process, and (ii) the \emph{reward mechanism}, which is the mechanism giving rewards to correct validators  that decided on the new block. Informally, if the selection mechanism is fair, then every process will become validator proportionally to its merit parameter; and if the reward mechanism is fair then for each height only the correct validators get a reward. By combining the two mechanisms, a correct process gets rewarded at least a number proportional to its merit parameter, since the faulty processes do not get any reward.
	
We define the following properties for characterizing the fairness of a reward mechanism. For each height, each validator has a boolean variable which we call a \emph{reward parameter}.
	\begin{enumerate}
		\item \label{fairness1} For each block in the blockchain, all  correct validators for that block have a reward parameter equal to $1$,
		\item \label{fairness2} For each block in the blockchain, all faulty validators and the processes that are not validators should have a reward parameter  for that block equal to $0$.
		\item \label{fairness4} A process gets a reward for a block if and only if it has a reward parameter for that block equal to $1$.
		\Pitem \label{fairness4bis} There exists a height $H$ such that for a block in the blockchain at   height $H' > H$ a process gets a reward for that block if and only if it has a reward parameter for that block equal to $1$.
	\end{enumerate}
	
	\setcounter{theorem}{2}
	\begin{definition}[Fairness of a reward mechanism]
		A reward mechanism is \emph{fair} if it satisfies the conditions \ref{fairness1}, \ref{fairness2} 
		and \ref{fairness4}.
	\end{definition}
	
	\begin{definition}[Eventual fairness of a reward mechanism]
		A reward mechanism is \emph{eventually fair} if it satisfies the conditions \ref{fairness1}, \ref{fairness2} 
		and \ref{fairness4bis}.
	\end{definition}
%

	\begin{definition}[(Eventual) Fairness of a blockchain protocol]
		A blockchain protocol is \emph{fair} (resp. \emph{eventually fair}) if it has a fair selection mechanism and a fair (resp. eventually fair) reward mechanism.
	\end{definition}
	
	
	

\subsection{Tendermint's Reward Mechanism}\label{ssec:tendermintReward}
The validators selection mechanism is part of a separate module of Cosmos project  \cite{cosmos}, which has Tendermint as core-blockchain. The selection mechanism is today left configurable by the application, therefore in the following we do not address this part. The  rewarding mechanism, on the other hand, referred as the \textit{Tendermint's reward mechanism}  is part of the original Tendermint protocol and it is reported in Figure \ref{fig:repConsensus} at lines \ref{Commit-wait} and \ref{ComDel-01} - \ref{ComDel-02}. Tendermint's reward mechanism works  as follows:
	\begin{itemize}
		\item Once a new block is decided for height $H$, processes wait for $\TimeOutCommit$ time to collect the decision from the other validators for $H$, and put them in their set $\toReward$ (Fig. \ref{fig:repConsensus}, lines \ref{Commit-wait} and \ref{ComDel-01} - \ref{ComDel-02}).
		\item During the consensus at height $H$, let us assume that $p_i$ proposes the block that will get decided in the consensus. $p_i$ proposes to  reward  processes in its set $\toReward$ (Fig. \ref{fig:repConsensus}, line \ref{repC:cons}). That is, only the processes from which $p_i$ delivered a commit will get a reward for the block at height $H-1$.

	\end{itemize}

\setcounter{theorem}{11}	
\begin{lemma}\label{l:tendermintNotFair}
	The reward mechanism of Tendermint is not eventually fair.
\end{lemma}

\begin{proofL}
	We assume that the system becomes synchronous, and that $\TimeOutCommit < \Delta$, where $\Delta$ is the maximum message delay in the network. 
	For any height $H$, let $p_i$ be a validator for the height $H-1$ and $p_j$ the validator whose proposal get decided for the height $H$.
	It may happen that $p_j$ did not receive the commit from $p_i$ before proposing its block.
	Hence when the block is decided, $p_i$ does not get a reward for its effort, which contradicts the condition \ref{fairness4bis} of the reward mechanism fairness.
	Tendermint's reward mechanism is not eventually fair.
	\renewcommand{\toto}{l:tendermintNotFair}
\end{proofL}

	Let us observe that to make Tendermint's reward mechanism at least eventually fair it is necessary to increase $\TimeOutCommit$ for each round until it catches up the message delay. We refer to this variant as the \textit{Tendermint's reward mechanism with modulable timeouts.} Moreover, the commit message should contain enough information to keep track of process participation in each phase, in order to exclude from the reward a process that did not send a propose or vote message but that sends a commit because he is aware about the block produced. This scenario can be avoided, for instance, by including in the $toReward$ variable a process $p_i$ only if $f+1$ commit messages contain the process $p_i$. 
	\begin{lemma}\label{l:tendermintModFair}
		When the commit messages are sufficient to detect a process that did not send expected messages,  Tendermint's reward mechanism with modulable timeouts is eventually fair.
	\end{lemma}
	\begin{proofL}
		We change the reward mechanism in Tendermint  as follows: 
		\begin{itemize}
			\item Once a new block is decided, say for height $H$, processes wait for at most \TimeOutCommit \ to collect the decision from the other validators for that height, and put them in their set $\toReward$.
			\item If a process did not get the commits from all the validators for that height before the expiration of the time-out, it increases the time-out for the next height.
			\item During the consensus at height $H$, let us assume that $p_i$ proposes the block that will get decided in the consensus. $p_i$ gives the reward to the processes in its $\toReward$. 
		\end{itemize}
		In this reward mechanism, $\TimeOutCommit$ is increased whenever a process does not have the time to deliver all the commits for the previous round.
		We prove that this reward mechanism is eventually fair.
		
		There is a point in time $t$ from when the system will become synchronous, and all the commits will be delivered by correct processes before the next height.
		From the time $t$, at height $H$ all correct processes know the exact set of validators that committed the block from $H-1$, and from those commit messages, they can exclude  the set of processes that did not participate to the consensus, and they give to those validators from $H-1$ a reward parameter greater than $0$.
		The validators in $H$  give the reward to the correct validators that committed and which are the only one with a reward parameter greater than $0$ for $H-1$, which satisfy the fairness conditions \ref{fairness1}, \ref{fairness2} 
		and \ref{fairness4bis}, so the reward mechanism presented is eventually fair.
		\renewcommand{\toto}{l:tendermintModFair}
	\end{proofL}
	
%

	
	\begin{theorem}\label{t:tendermintFair}
		In an eventual synchronous system, if the selection mechanism is fair, and from the commit messages processes can differentiate correct and Byzantine behaviour, then Tendermint Repeated Consensus with the reward mechanism with modulable time-outs, is eventually fair.
	\end{theorem}
	\begin{proofT}
		The proof follows by Lemma \ref{l:tendermintModFair}.
		\renewcommand{\toto}{t:tendermintFair}
	\end{proofT}
 
\subsection{Necessary and Sufficient Conditions for a Fair Reward Mechanism}\label{ssec:resultFairness}
	In this section, we discuss the consequences of the synchrony on the existence of a fair reward in Repeated Consensus based  blockchain protocols where  the blockchain is constructed by a mechanism of repeated consensus. That is,  at each height, a subset of processes called validators produce a block executing an instance of One-Shot Consensus.
	
	\begin{theorem}\label{t:strongFairness}
		There exists a fair reward mechanism in a Repeated Consensus based blockchain protocol iff the system is synchronous.
	\end{theorem}
	\begin{proofT}
		We prove this theorem by double implication.
		\begin{itemize}
			\item If the system is synchronous, then there exists a fair reward mechanism.
				
				We assume that the system is synchronous and all messages are delivered before the $x$ following blocks. We consider the following reward mechanism.
				For all the correct validators at any height $H$, if $H-x \le 0$, do not reward yet, otherwise:
				\begin{itemize}
					\item Set to $1$ the reward parameters of all correct validators in $H-x$, and to $0$ the merit parameters of the others.
					\item Reward only the validators with a reward parameter equal to $1$ from the height $H-x$.
				\end{itemize}
				We prove in the following that the above reward mechanism is fair.
				
				Note that the system is synchronous and messages sent are delivered within at most $x$ blocks.
				Therefore, the exact set of correct validators at $H-x$ is known by all at $H$. 
				By construction, the validators in $H$ exactly give the reward to the correct validators who are the only one with a reward parameter for $H-x$ greater than $0$, which satisfies the fairness conditions (\ref{fairness1}, \ref{fairness2} 
				and \ref{fairness4}).
				
				
			\item If there exists a fair reward mechanism, then the system is synchronous.
				
				By contradiction, we assume that $\mathcal{P}$ is a protocol having a fair reward mechanism and that, the system is not synchronous.
				We say that the validators following $\mathcal{P}$ are the correct validators.
				Let $V^{i}$ be a set of validators for the height $i$, and $V^{j}, (j >i)$ be the set of validators who gave the reward to the correct validators in $V^i$. 
				Since the system is not synchronous, the validators in $V^{j}$ may not receive all messages from $V^{i}$ before giving the reward.
				
				By conditions \ref{fairness1}, and \ref{fairness2}, 
				it follows that all and only the correct validators in $V^i$ have a reward parameter greater than $0$.
				Since the reward mechanism is fair, with the condition \ref{fairness4}, we have the validators in $V^{j}$ gave the reward only to the correct validators in $V^i$. That means that the correct validators in $V^j$ know exactly who were the correct validators in $V^i$,
				so they got all the messages before giving the reward. Contradiction, which conclude the proof.
		\end{itemize}
		\renewcommand{\toto}{t:strongFairness}
	\end{proofT}
	
	If there is no synchrony, then there cannot be a fair consensus based protocol for blockchain. 
	The fairness we define states that every time during the execution, the system is fair, so if a process leaves the system, it receives all rewards it deserves for the time it was in the system.
	
	\begin{theorem}\label{t:eventualFairness}
		There exists an eventual fair reward mechanism in a Repeated Consensus based blockchain protocol iff the system is eventually synchronous or synchronous.
	\end{theorem}
	\begin{proofT}
		We proof this result by double implication.
		\begin{itemize}
			\item If the system is eventually synchronous or synchronous, then there exists an eventual fair reward mechanism.
				
				If the system is synchronous, the proof follows directly from theorem \ref{t:strongFairness}. Otherwise, we prove that the following reward mechanism is eventually fair.
				When starting the height $H$, the correct validator for $H$ do the following:
				\begin{itemize}
					\item Start the time-out for the reception of the messages from validators for the block $H-1$;
					\item Wait for receiving the messages from validators for the block $H-1$ or the time-out to expire.
						If the time-out expires before the reception of more that $2n/3$ of the messages, increase the time-out for the next time;
					\item Set to $1$ the reward parameters for $H-1$ of the correct validators which messages where received, and to $0$ the merit parameters of the others;
					\item Reward only the validators for the height $H-1$ which have a merit parameter different than $0$.
				\end{itemize}
				Since the system is eventually synchronous, eventually when the system will become synchronous, the processes, in particular the validators for $H$ will receive messages for all the correct validators from round $H-1$ before the round $H$.
				We note that the condition \ref{fairness4bis} is a weaker form of the condition \ref{fairness4} where we do not consider the beginning.
				So we end the proof by applying the theorem \ref{t:strongFairness} from the time when the system becomes synchronous.
				
			\item If there exists an eventual fair reward mechanism, then system is eventually synchronous or synchronous.
				
				If the reward mechanism is fair, by theorem \ref{t:strongFairness}, the communication is synchronous, which ends the proof.
				Otherwise, since the reward mechanism is eventually fair, then there is a point in time $t$ from when all the rewards are correctly distributed.
				By considering $t$ as the beginning of our execution, then we have that the reward mechanism is fair after $t$, so by the theorem \ref{t:strongFairness}, the system is synchronous from $t$.
				If the system were not synchronous before $t$, that means that it is eventually synchronous, otherwise, it is synchronous.
				Which ends the proof.
		\end{itemize}
		\renewcommand{\toto}{t:eventualFairness}
	\end{proofT}
 
 	\begin{corollary}\label{c:asyncFairness}
		In an asynchronous system, there is no (eventual) fair reward mechanism, so if the communication system is asynchronous, then there is no (eventual) fair Repeated Consensus-based Blockchain protocol.
		Note that this result is valid even if all the processes are correct.
	\end{corollary}

	
	\section{Conclusion} \label{sec:conclusion}

The first contribution of this  paper is the improvement and the formal analysis of the original Tendermint  protocol, a PBFT-based repeated consensus protocol where the set of validators is dynamic.  Each improvement we introduce is motivated by bugs  we discover in the original protocol. 
A preliminary version of this paper has been reported in  \cite{ADPT18}. 
Very recently a new version of Tendermint  has been advertised  in \cite{BKM18} by Tendermint foundation without an operational release. The authors argue that their solution works if  the two hypothesis below are verified: \emph{Hypothesis 1}: if a correct process  receives some message $m$ at time $t$, all correct processes will receive $m$ before $max(t, global \ stabilization \ time) + \Delta$. Note that this property called by the authors \emph{gossip communication} should be verified even though $m$ has been sent by a Byzantine process. \emph{Hypothesis 2}: there exists eventually a proposer such that its proposed value will be accepted by all the other correct processes. 
Moreover, the formal and complete correctness proof of this new protocol is still an open issue (several not trivial bugs have been reported recently e.g. \cite{issue36-37}).


Our second major contribution is the study of the fairness of the reward mechanism in repeated-consensus blockchains. We proved that  there exists a reward mechanism in repeated-consensus blockchains that  is (eventually) fair if and only if the system communication is (eventually) synchronous.  
In addition, we show that even if Tendermint protocol evolves in an eventual synchronous setting, it is not eventually fair. 
However,  it becomes eventually fair when timeouts are carefully tuned, and under the assumption that commit messages contains enough information to distinguish between correct and Byzantine processes in the synchronous period. Our study opens  an interesting  future research direction related to the fairness of the selection mechanism in repeated-consensus based blockchains.


	\newpage
	\bibliographystyle{plain}
	\bibliography{biblio,mariapotop,hypercoin}
%


\end{document}